\documentclass{aa}  

\usepackage{graphicx}

\usepackage{epstopdf}
\usepackage{txfonts}
\usepackage{amsmath}    
\usepackage[dvipsnames]{xcolor}

\begin{document}

   \title{Shell helium-burning hot subdwarf B stars as candidates for blue large-amplitude pulsators}


   \author{H. Xiong,
          \inst{1,2,3}
          \and
          L. Casagrande,\inst{2}
         \and
         X. Chen,\inst{1,3,4} 
         \and
         J. Vos,\inst{5}
         \and
         X. Zhang,\inst{6}
         \and
         S. Justham,\inst{7,8,9,10,11}
         \and
         J. Li,\inst{13}
         \and
         T. Wu,\inst{1,3,4,12}
         \and
         Y. Li,\inst{1,3,4}
         \and
         Z. Han,\inst{1,3,4}
         }

   \institute{Yunnan Observatories, Chinese Academy of Sciences, 396 Yangfangwang, Guandu District, Kunming, 650216, P. R. China  \label{inst1}          
         \and
             Research School of Astronomy and Astrophysics, The Australian National University, Canberra, ACT 2611, Australia \label{inst2}      \\
             \email{heran.xiong@anu.edu.au}
          \and
          Key Laboratory for the Structure and Evolution of Celestial Objects, Chinese Academy of Sciences, 396 Yangfangwang, Guandu District, Kunming, 650216, P. R. China\label{inst3} 
          \and
          Center for Astronomical Mega-Science, Chinese Academy of Sciences, 20A Datun Road, Chaoyang District, Beijing, 100012, P. R. China\label{inst4} 
          \and
          Astronomical Institute of the Czech Academy of Sciences, CZ-25165, Ondřejov, Czech Republic\label{inst5} 
          \and
          Department of Astronomy, Beijing Normal University, Beijing, 100875, P. R. China\label{inst6} 
          \and
          School of Astronomy \& Space Science, University of the Chinese Academy of Sciences, Beijing 100012, China\label{inst7} 
          \and
          The Key Laboratory of Optical Astronomy, National Astronomical Observatories, The Chinese Academy of Sciences, Datun Road, Beijing 100012, China\label{inst8} 
          \and
          University of Chinese of Academy of Science, Yuquan Road 19, Shijingshan Block, Beijing 100049, P. R. China\label{inst9} 
          \and
          Anton Pannekoek Institute for Astronomy, University of Amsterdam, Science Park 904, 1098XH Amsterdam, The Netherlands\label{inst10} 
          \and
          Max Planck Institute for Astrophysics, Karl-Schwarzschild-Str. 1, 85748 Garching, Germany\label{inst11}
          \and
          Institute of Theoretical Physics, Shanxi University, Taiyuan 030006, China\label{inst12}
          \and 
          Key Laboratory of Space Astronomy and Technology, National Astronomical Observatories, Chinese Academy of Sciences, Beijing 100101, People's Republic of China\label{inst13}
                          }

   \date{Received September 15, 1996; accepted March 16, 1997}

 
  \abstract
  {Blue large-amplitude pulsators (BLAPs) are a newly discovered type of variable star. 
Their typical pulsation periods are on the order of a few tens of minutes, with relatively large amplitudes of 0.2-0.4 mag in optical bands, and their rates of period changes are on the order of $10^{-7} yr^{-1}$ (both positive and negative).  
They are extremely rare objects and attempts to explain their origins and internal structures have attracted a great deal of attention. 
Previous studies have proposed that BLAPs may be pre-white dwarfs, with masses around $0.3M_\odot$, or core-helium-burning stars in the range of $\sim 0.7-1.1M_\odot$.
In this work, we use a number of MESA models to compute and explore whether BLAPs could be explained as shell helium-burning subdwarfs type B (SHeB sdBs). 
The models that best match existing observational constraints have helium core masses in the range of $\sim 0.45-0.5M_\odot$.
Our model predicts that the positive rate of period change may evolve to negative.
The formation channels for SHeB sdBs involve binary evolution and although the vast majority of BLAPs do not appear to be binaries (with the exception of HD 133729), the observational constraints are still very poor. Motivated by these findings, we explored the Roche lobe overflow channel. Of the 304 binary evolution models we computed, about half of them are able to produce SHeB sdBs in long-period binaries that evade detection from the limited observations that are currently available. }

   \keywords{(Stars:) binaries: general --
             Stars: oscillations (including pulsations) --
             Stars: peculiar (except chemically peculiar) --
             (Stars:) subdwarfs
               }

   \maketitle
%

\section{Introduction}
Pulsating stars are important in astrophysics. 
For example, through asteroseismology, it is possible to infer the internal structure of pulsating stars 
and use this information to calibrate some basic parameters of stellar physics \citep{Aerts,Chaplin,Bedding,Balona11,Huber2012,NatAst}. \cite{Pietrukowicz} discovered a new class of pulsators, with periods of 20 to 40 min and large amplitudes, namely, from 0.2 to 0.4 mags in the optical. By comparison, other pulsators with similar periods, for example, rapidly oscillating Ap stars pulsating in the range of 5 to 24 min have amplitudes from 0.001 to 0.02 mag \citep[e.g., ][]{hold}. 

Moreover, this new class of pulsators has sawtooth-shaped light curves,
similar to those of the fundamental radial mode pulsations of classical Cepheids and RR Lyrae-type stars, 
and effective temperatures of about 30,000K,
similarly to hot subdwarf O/B stars (sdOB), but with much larger luminosities or lower surface gravities 
that is ${\rm log} g {\rm (g/cm s^{-2})} < 5.3$. 
Long-term observations show that the pulsations are very stable and their rates of period change ($r=\dot{P}/P$) are on the order of $10^{-7}\rm yr^{-1}$). \citet{Pietrukowicz} named these stars "blue large-amplitude pulsators \citep[BLAPs; see for example][for a succinct review of their properties]{cor2019}. 

Such stars are extremely rare: among nearly half a million pulsating stars, the OGLE survey \citep{Pietrukowicz,Soszy2013,Soszy2014,Soszy2015,Soszy2015a,Soszy2016,mr} found only 14 BLAPs in the Galaxy and none in the Magellanic System \citep{Pietrukowicz2018}.  
\cite{Ramsay} investigated 10 of the 14 known BLAPs with parallaxes available from Gaia DR2.
By dereddening their colors,
 \cite{Ramsay} found that 6 of them  had absolute magnitudes and intrinsic colors
consistent with the temperature derived from optical spectra in \cite{Pietrukowicz} 
as well as from the theoretical predictions of \cite{By}, thus confirming their  nature as BLAPs. The remaining four were redder and fainter, therefore suggesting that they could have been other types of pulsating variables.

Using Zwicky Transient Facility (ZTF), \citet{Kupfer2019} discovered another class of pulsators, with amplitudes similar to those of BLAPs, but with shorter periods and higher surface gravities and effective temperatures. Therefore, \citet{Kupfer2019} named them high-gravity BLAPs. Moreover, their rarity is comparable that of BLAPs.
Recently, \citet{McWhirter} crossmatched Gaia DR2 with the ZTF DR3 and analyzed the period-folded light curves. They identified another 16 BLAPs candidates and six high-gravity BLAPs candidates from over 162 million sources. Many of these candidates actually have pulsation periods longer than the BLAPs originally discovered by \cite{Pietrukowicz}. Thus, it is likely that the pulsation periods for BLAPs span from a few minutes to almost one hour. \citet{Lin} comprehensively studied a peculiar BLAP, TMTS-BLAP-1 aka ZGP-BLAP-01, and gave more detailed physical properties.

In Table 1, we list some parameters for confirmed and candidate BLAPs and high-gravity BLAPs: pulsation period ($P$), its change rate ($r=\dot{P}/P$),
effective temperature ($T_{\rm eff}$), 
surface gravity (${\rm log} g$), and helium to hydrogen number ratio ($\log{N_{\rm He}/N_{\rm H}}$). 
The symbols $\surd$ (or $\ast$) indicates whether absolute magnitudes and intrinsic colors from \cite{Ramsay} are consistent (or inconsistent) with the effective temperatures from the literature; the symbols $\Uparrow$ (or $\odot$) denotes confirmed BLAPs (or high-confidence BLAPs).
From this table, we see that pulsation periods are in the range of $17-60 \rm{mins}$ (excluding high-gravity BLAPs), that is, significantly larger than those initially discovered. The high-gravity BLAPs have periods in the range of $2-8 \rm{mins}$. It is worth noticing the apparent existence of a period gap between BLAPs and high-gravity BLAPs. 

\begin{table*}
\caption{List of confirmed and candidate BLAPs and high-gravity BLAPs with some of their parameters from the discovery papers.  
Symbols $\surd$ ($\ast$) in the status column denote that their absolute magnitudes and intrinsic colors dereddended by the parallax from Gaia DR2 data are consistent (inconsistent) with the temperatures derived from optical spectra or theoretical predictions \citep[see][]{Ramsay}. For BLAPs from Ref (3), the symbols $\Uparrow$ denote confirmed status, and $\odot$ denote high-confidence candidates.} 
 \label{table 1}
\centering
\begin{tabular}{lccccccc}     
\hline\hline
 Variable & $P({\rm min})$& $\grave{P}/P(10^{-7}\rm yr^{-1})$ &$T_{\rm eff}$  &$\log{g}$&$\log{N_{\rm He}/N_{\rm H}}$& status & Ref. \\
 \hline
   OGLE-BLAP-001 &28.26 &2.90$\pm$3.70 &30800$\pm$500&4.61$\pm$0.07 &-0.55$\pm$0.05&$\surd$ & (1) \\
   OGLE-BLAP-002  &23.29 &-19.23$\pm$8.05  & -&- &-&  & (1) \\
   OGLE-BLAP-003  &28.46 &0.82$\pm$0.32 &- &- & -&$\ast$ & (1) \\
   OGLE-BLAP-004  &22.36 &-5.03$\pm$1.57 &- & -& -& $\ast$ & (1) \\
   OGLE-BLAP-005  &27.25 &0.63$\pm$0.26 &- &- &-&  & (1) \\
   OGLE-BLAP-006  &38.02 &-2.85$\pm$0.31 &- &- &- &$\ast$ & (1) \\
   OGLE-BLAP-007  &35.18 &-2.40$\pm$0.51 &- &- &- & & (1) \\
   OGLE-BLAP-008  &34.48 &2.11$\pm$0.27 &- &- &-&  & (1) \\
   OGLE-BLAP-009  &31.94 &1.63$\pm$0.08 &31800$\pm$1400&4.40$\pm$0.18 &-0.41$\pm$0.13&$\surd$ & (1) \\
   OGLE-BLAP-010  &32.13 &0.44$\pm$0.21 &- &- &-& $\surd$ & (1) \\
   OGLE-BLAP-011  &34.87 &6.77$\pm$8.87 &26200$\pm$2900&4.20$\pm$0.20 &-0.45$\pm$0.11&$\surd$ & (1) \\
   OGLE-BLAP-012  &30.90 &0.03$\pm$0.15 &- &- &-& $\surd$ & (1) \\
   OGLE-BLAP-013  &39.33 &7.65$\pm$0.67 &- &- &-& $\ast$ & (1) \\
   OGLE-BLAP-014  &33.62 &4.82$\pm$0.39 &30900$\pm$2100&4.42$\pm$0.26 &-0.54$\pm$0.16&$\surd$ & (1) \\
   high-gravity-BLAP-1& 3.34& -& 34000$\pm$500& 5.70$\pm$0.05&- & & (2) \\
   high-gravity-BLAP-2& 6.05& -& 31400$\pm$600& 5.41$\pm$0.06& -& & (2) \\
   high-gravity-BLAP-3& 7.31& -& 31600$\pm$600& 5.33$\pm$0.05& -& & (2) \\
   high-gravity-BLAP-4& 7.92& -& 31700$\pm$500& 5.31$\pm$0.05& -& & (2) \\
   ZGP-BLAP-01/TMTS-BLAP-1 	&18.933 & 22.3$\pm$0.9&  -& -& -&  & (3) (5)  \\
   ZGP-BLAP-02  &48.258 & -&  -& -& -&  & (3)  \\ 
   ZGP-BLAP-03  &53.705 & -&  -& -& -&  & (3)  \\
   ZGP-BLAP-04  &46.681 & -&  -& -& -&  & (3)  \\ 
   ZGP-BLAP-05  &54.000 & -&  -& -& -&    $\odot$& (3)\\ 
   ZGP-BLAP-06  &35.839 & -&  -& -& -&    $\odot$& (3)\\
   ZGP-BLAP-07  &44.627 & -&  -& -& -&    $\odot$& (3)\\                 
   ZGP-BLAP-08  &35.137 & -&  -& -& -&    $\odot$& (3)\\ 
   ZGP-BLAP-09  &23.264 & -&  -& -& -&    $\Uparrow$& (3)\\ 
   ZGP-BLAP-10  &55.180 & -&  -& -& -&    $\odot$& (3)\\ 
   ZGP-BLAP-11  &27.951 & -&  -& -& -&    $\odot$& (3)\\  
   ZGP-BLAP-12  &51.619 & -&  -& -& -&    & (3)\\ 
   ZGP-BLAP-13  &21.578 & -&  -& -& -&    & (3)\\  
   ZGP-BLAP-14  &17.016 & -&  -& -& -&    $\odot$& (3)\\ 
   ZGP-BLAP-15  &51.073 & -&  -& -& -&    & (3)\\ 
   ZGP-BLAP-16  &37.330 & -&  -& -& -&    & (3)\\ 
   ZGP-HGBLAP-01    &2.428 & -&  -& -& -&    & (3)\\
   ZGP-HGBLAP-02        &6.071 & -&  -& -& -&    $\odot$& (3)\\ 
   ZGP-HGBLAP-03        &4.203 & -&  -& -& -&    $\odot$& (3)\\  
   ZGP-HGBLAP-04        &5.950 & -&  -& -& -&    $\odot$& (3)\\  
   ZGP-HGBLAP-05        &8.240 & -&  -& -& -&    & (3)\\ 
   ZGP-HGBLAP-06        &6.260 & -&  -& -& -&    $\odot$& (3)\\  
   HD 133729  &32.27 &-11.5 & 29000 & 4.5 & -& & (4)\\
 \hline\hline
\end{tabular}
\begin{minipage}{1\textwidth}
(1) \cite{Pietrukowicz}, (2) \cite{Kupfer2019}, (3) \cite{McWhirter}, (4) \cite{Pigulski}, (5) \cite{Lin}. The $\dot{P}/P$ of OGLE-BLAP-002,-004,-011 are from \cite{Wu} and it of ZGP-BLAP-01/TMTS-BLAP-1 is from \cite{Lin}.
\end{minipage}
\end{table*}


In the original paper, \cite{Pietrukowicz} proposed two possible models accounting for the characteristics of BLAPs that is they are proto low-mass white dwarfs (pre-WD model, $\sim 0.3M_\odot$) or core-helium-burning stars with hydrogen-rich inflated envelopes ($\sim 1.0M_\odot$, CHeB hot subdwarfs model). 
Based on the two models, several researchers have tried to reproduce the pulsational properties of BLAPs.
\cite{Romero} and \cite{cor} simulated various pre-WD models and
provided two possible interpretations for the pulsation modes to account for the pulsation period, that is, the radial fundamental mode and high-order nonradial g modes.
However, the rate of period change from radial fundamental mode is higher than observed. 
The high-order nonradial g modes have comparable rates of period change 
but are inconsistent with the observation that only a single period is detected in BLAPs.
\cite{Wu} obtained similar results for the pre-WD model. 
In their study, pre-WDs with mass around $0.36 M_\odot$ have properties (effective temperature, surface gravity, and pulsation period from radial fundamental mode) similar to those of BLAPs, except for the rate of period change, that is, $r\sim10^{-5}\rm yr^{-1}$,  much higher than what has been observed. 

\cite{Wu} also investigated CHeB hot subdwarf models in detail and found that such stars with masses of 0.7–1.1$M_\odot$ are in good agreement with almost all BLAPs properties. 
Their models have proper surface helium-to-hydrogen number ratios and rates of period change ($r\sim 10^{-7}\rm yr^{-1}$) in addition to effective temperatures, surface gravities, and pulsation periods.  
Based on their study, BLAPs are objects in the middle to late phases of core helium burning stage, with metal-rich models matching $r$ better than metal-poor ones. 
In their models, however, the H-rich envelope mass is fixed and determined by the total mass. Binary evolution in fact will produce hot subdwarfs with similar total mass but various H-rich envelopes and the evolution properties remarkably depend on the envelope mass \citep{Han2002,Han2003,Heber2009,Heber2016,Xiong}.  

On the other hand, using nonadiabatic analysis with the GYRE stellar oscillation code, \citet{Byrne} investigated the pulsation properties of the pre-WD model and the driving mechanism.
They showed that, if effects of atomic diffusion and radiative levitation were included,
the opacity bump at the iron opacity peak leads to a large instability region, that is, effective temperature from 30 000 K up to 50 000 K at least, ${\rm log} g$ from 3 to 7, and the periods of unstable fundamental modes from around 100 seconds up to 2-3 hours.
This range encompasses both BLAPs and high-gravity BLAPs.
\citet{Byrne2021} then further studied formation channels for BLAPs based on the pre-WD model, using a binary population synthesis method, and showed that both common envelope ejection (CEE) and stable Roche lobe overflow (RLOF) can produce these objects. However, CEE generally produces binaries with short orbital periods. The fact that until recently no companions have been detected for BLAPs \citep{Pietrukowicz,Kupfer,Ramsay,McWhirter2020} seems inconsistent with this channel.
However, one object worthy of attention in Table 1 is HD 133729. 
\citet{Pigulski} reported a BLAP orbiting the main-sequence B-type star HD 133729 on an orbital period of 23.08433 d.  
The pulsation period of this BLAP is 32.37 min, with an amplitude 0.21 mag 
and a rate of period change of $-11.5\times10^{-7}\rm {yr^{-1}}$, 
which is consistent with that of known BLAPs. 
The BLAP nature of the companion to HD 133729 was previously missed 
due to the dilution of the observed amplitude by the brighter primary. 
The companion of HD 133729 is the only BLAP in a binary system that is presently known, 
but the discovery of this object suggests that (at least some other) BLAPs might be hidden in binaries which might have been misclassified or missed for various reasons.

Since the origin and structure of BLAPs is still an open question and they have  temperatures similar to hot subdwarfs but with inflated envelopes, here we propose that BLAPs are shell-helium-burning (SHeB) hot subdwarfs. 
In the following, we systematically investigate the observational properties of such models based on a simple asteroseismic analysis and we study the parameter space for producing such objects via binary evolution. Our models are able to aptly replicate existing observations of BLAPs.  The paper is organized as follows. First, we describe our construction of certain sdB models, which possess different combinations of H-rich envelope mass and He core mass, and we investigate their period, $P,$ and relative pulsation rate of period change, $\dot{P}/P,$ along the evolutionary tracks. We then analyze how sdBs can form through a RLOF channel in Section 3, along with estimations of their number in the Galaxy and some of the observational signatures of putative companions. We present our conclusions in Section 4. 

\section{Configuration of BLAPs}
Subdwarf B-type stars (hereafter sdBs) are core-helium-burning stars (CHeB) with thin H-rich envelopes, located at the extreme horizontal branch (EHB) of the Hertzsprung Russell diagram \citep[e.g., ][]{Heber2009,Heber2016,Geier}.
The shell helium-burning (SHeB) follows the CHeB phase 
and we show in this section that 
some SHeB sdBs may aptly reproduce the properties of BLAPs. 

We constructed a certain number of sdB models and study the pulsation periods and their changes along evolutionary tracks.
The sdB models are computed with the stellar evolution code named MESA  \citep[Modules for Experiments in Stellar Astrophysics, version 15140, see][]{Paxton2011,Paxton2013,Paxton2015,Paxton2018,Paxton2019}.
Our study is for Population I stars (i.e., with a metallicity $Z=0.02$, as in \citealt{Pietrukowicz}). We adopted the nuclear network
$pp\underline{\makebox[1em]{}}cno\underline{\makebox[1em]{}}
extras\underline{\makebox[1em]{}}o18\underline{\makebox[1em]{}}ne22.net$,
which includes all relevant reactions for H and He burning, 
and the opacity table OPAL type II, which allows the abundances of C and O to vary with time.
The mixing length parameter, $\alpha_{\rm{MLT}}$, is set to 2. 
For simplicity, no stellar wind or other mass loss mechanism is included in our calculations. 
The input physics used here is similar to that of previous studies of Population I sdB stars by \citep[][see also \citealt{Schindler}]{Xiong}.

\begin{figure*}
  \centering
\includegraphics[scale=0.3]{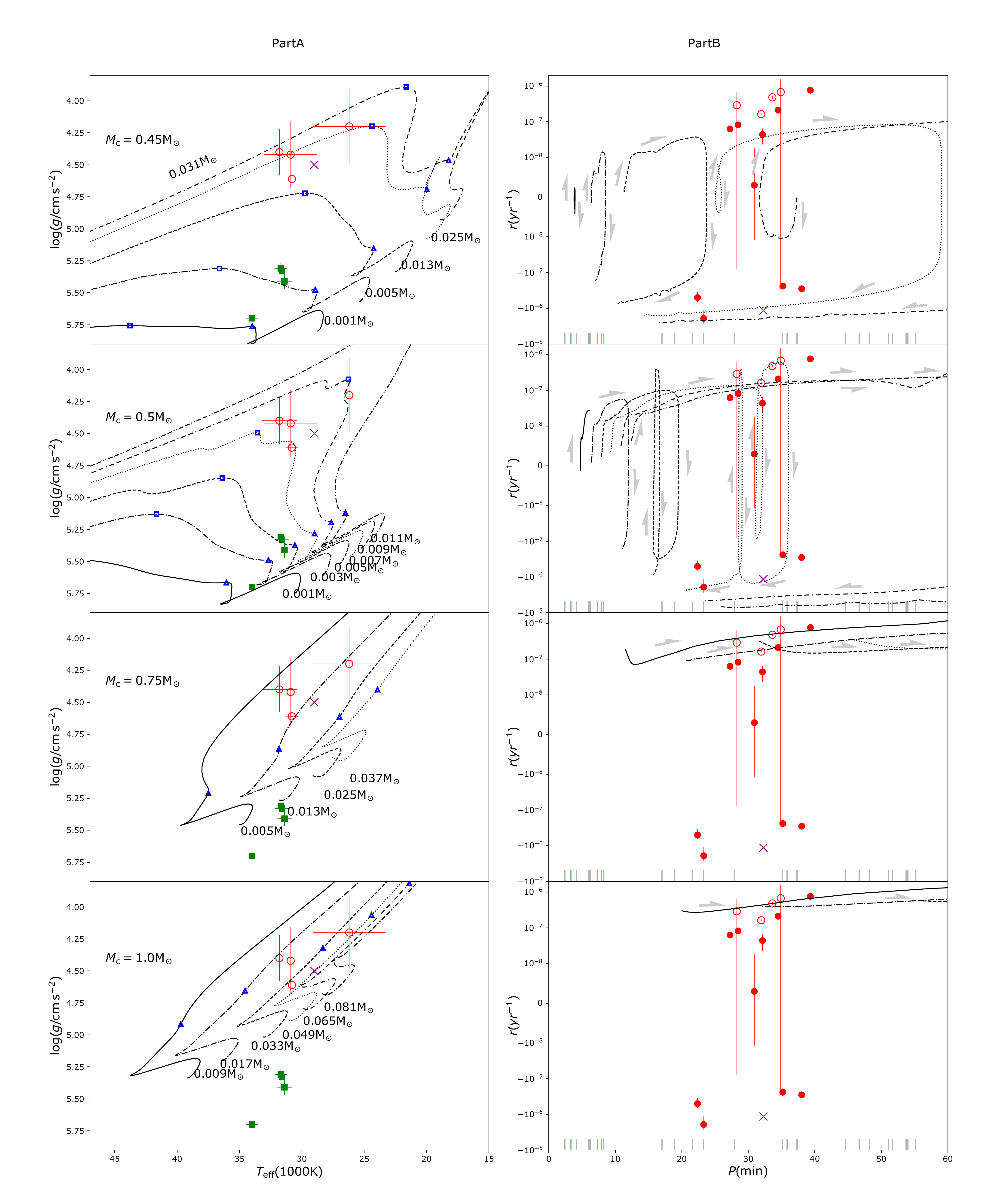}
\caption{Evolutionary tracks for constructed sdBs on the
  ($T_{\rm eff}-{\rm log} g$) diagram (left). In each panel the core mass
  $M_{\rm{c}}$ is reported, along with different values of H-rich envelope mass,
  $M_{\rm{e}}$, indicated with different linestyles. Four BLAPs (red) and four
  high-gravity BLAPs (green) with known $T_{\rm eff}$ and ${\rm log} g$ are
  overplotted with error bars for comparison (see Table 1).  In the top panel,
  the loops on the tracks with $M_{\rm e}=0.025M_\odot$ and $0.031M_\odot$ are
  caused by helium core breathing pulses \citep{Li2012,Li2017,Li}. To the
  right, we show the rate of period change $r=\grave{P}/P$ versus pulsation
  period $P$ for SHeB sdBs, for the same evolutionary tracks of the left panel.
  The observations of BLAPs \citep[with $r$]{Pietrukowicz}, high-gravity BLAPs
  \citep[without $r$]{Kupfer} and the candidates \citep[without $r$]{McWhirter}
  are shown respectively with red dots (open circles highlighting stars with
  spectroscopic parameters), green ticks (shown as green squares on the left
  panel), and grey ticks (not shown on the left panel). Purple $\times$
  indicates HD 133729 on all panels. There are some markers on both panels to
  indicate special evolutionary points, and the same markers on the left and
  right panels are for the same point. $\Delta$ indicates the starts of stable
  SHeB, and $\Box$ indicates where are the maximum radius of SHeB phases. So
  SHeB sdBs evolve from $\Delta$ to $\Box$. When the core masses are small
  ($0.45$ or $0.5 M_\odot$), the SHeB sdBs expand to $\Box$ and then shrink to
  WD ($\Delta \rightarrow \Box \rightarrow outside$ ); when the core masses are
  large ($0.75$ or $1.0 M_\odot$), the SHeB sdBs directly expand to giant branch
  ($\Delta \rightarrow outside$). The arrows on right panels also indicate the
  evolutionary direction. We note that the r-axis on Part B applies symlog
  (symmetrical log) scale, which allows positive and negative values by setting
  a range around zero within the plot to be linear instead of logarithmic.}
\label{fig combine}
\end{figure*}

\subsection{The sdB model}\label{sec:sdBmodel}
The main characteristics of sdBs are determined by the helium core mass, $M_{\rm c}$, 
the H-rich envelope mass, $M_{\rm e}$, and the element abundances in the envelope. 
We start by constructing some zero-age sdBs, that is at the onset of helium burning quietly in the center. 
We adopt four core masses that is $M_{\rm c}$=0.45, 0.5, 0.75, and 1.0 $M_\odot$, respectively\footnote{Since we are not interested in the ignition process, nor do we follow the evolution of a star before reaching the sdB phase, in this work we start our analysis from helium-burning quietly in the core.}.
For each core mass, we explore a series of envelope masses starting from an initial value of $M_{\rm e}=0.001M_\odot$ and increasing it in steps of $0.002M_\odot$, until the evolutionary tracks are beyond the region in the $T_{\rm eff}-\log{g}$ space where BLAPs are found. In total, we computed more than 80 models, some of which are displayed in Figure \ref{fig combine}. 
In the core, the adopted helium abundance is $Y$=0.98, 
while the number ratio of helium to hydrogen, $\log{N_{\rm He}/N_{\rm H}}$, in the envelope is set to -0.55 for simplicity. The latter value corresponds to the best atmosphere model fit for OGLE-BLAP-001, which is the prototype object of BLAPs \citep[see Table 1 and ][]{Pietrukowicz}.

We then evolved these sdBs to reach the SHeB phase and continue far beyond the location of BLAPs on the Kiel diagram ($T_{\rm eff} > 50\,000 \rm~K$ or $\log{g}<4.0$). 
The evolutionary tracks on the $T_{\rm eff}-{\rm log} g$ plane are presented on the left panels of Fig.\,\ref{fig combine}, 
where the legend indicates the core and envelope mass, and different line-styles mark models with the same envelope mass on the left and right panels.
For clarity, not all models are shown in this figure.
For comparison, four BLAPs (in red) and four high-gravity BLAPs (in green) with known $T_{\rm eff}$ and ${\rm log} g$ are plotted with their error bars (see Table 1), while the purple $\times$ indicates HD 133729.
For $M_{\rm c}=0.45M_\odot$ (upper panel), 
we see several loops on the evolutionary tracks when $M_{\rm e}=0.025M_\odot$ and $0.031M_\odot$, which are caused by helium core breathing pulses after the exhaustion of the central He \citep{Caloi,Li2012,Li2017,Li,Ostrowski}.

 \subsection{Pulsations and rate of period change, $\dot{P}/ P$ }

The large amplitude, short periods, and small rates of period change observed in BLAPs still pose a challenge to pulsation theory \citep{Pietrukowicz}. A few pioneering asteroseismic investigations have been carried out \citep[for example][]{Romero,cor,Wu,Byrne}, but none of them can fully explain the existence and the pulsation properties of BLAPs. 

In this paper, rather than focusing on a detailed asteroseismic analysis, we explore a new evolutionary pathway to BLAPs. Our goal is thus to compute a broad range of evolutionary models and use these to inform on possible formation channels for BLAPs via binary evolution. While we defer a detailed asteroseismic analysis to a future paper (Wu et al., in prep.), we use the fact that the the lightcurves of BLAPs are similar to those of classical pulsators like Cepheid and RR Lyrae-type stars that exhibit just the radial fundamental mode \citep[see for example][for a discussion of pulsation modes]{cor2019}. Thanks to this similarity, we can use simple scaling relations to estimate the pulsation period, $P,$ and rate of period change, $r=\dot{P}/ P,$ for our models. We describe our method further below. 

Confidence in this approach comes from the fact that $P$ and $r$ inferred from scaling relations agree to within 10 percent (5 percent for most BLAPs models) with those computed with detailed asteroseismic analysis via GYRE (Wu et al., in prep). This suffices for the qualitative investigation of Fig. \ref{fig combine}. Also, despite our here on SHeB sdB models, we also constructed some $\sim 0.3M_\odot$
pre-WD models; the values of $r$ we derive when these models are located in the BLAPs region of the HR diagram are on the order of $10^{-5}$, similar to those obtained by \cite{Wu} through detailed asteroseismic analysis.

For p-mode oscillations, we used Eq. (1) to estimate the frequency, ($\nu$), with radial order, $n,$ and spherical harmonic degree, $l$ \citep{Tassoul1980}, which is not dependent on the driving mechanism
 \begin{equation}
\nu (n,l) \approx (n+l/2+\epsilon) \Delta \nu,
\end{equation}
where $\epsilon$ is the phase constant ($=2.625$ according to the best-fitting model in \citealt{Wu}) and
$\Delta \nu$ is the large frequency separation, which is defined as the frequency spacing 
between adjacent radial order mode with the same spherical harmonic degree. The value of 
$\Delta \nu$ strongly depends on mean density and is estimated by \citet{Tassoul1980}
\begin{equation}
\Delta \nu \approx \Delta\nu_\odot\sqrt{\overline{\rho}/\overline{\rho_\odot}} =\Delta\nu_\odot \sqrt{\frac{M}{M_\odot}\left(\frac{R_\odot}{R}\right)^3},
\end{equation}
where $\overline{\rho}$, $M$, and $R$ are the mean density, mass, and radius of our sdBs models, respectively. $\nu (0,0)\approx \epsilon \Delta \nu$ is the frequency of fundamental mode, and $\Delta \nu_\odot = 135.1\mu {\rm HZ}$ \citep{Huber2011}. The period is the reciprocal of the frequency, $P=1/\nu(0,0)$.
Then, following the definition of \citet{Pietrukowicz} the rate of period change can be estimated as 
\begin{equation}
    r=\frac{\Delta P}{\Delta t}\frac{1}{P}\\
,\end{equation}
which we calculate in this work as:
\begin{equation}
    r_{\rm n}=\frac{P_{\rm n+1}-P_{\rm n}}{t_{\rm n+1}-t_{\rm n}}\frac{1}{P_{\rm n+1}}
,\end{equation}
where $n$ refers to each point along a track computed through MESA. In the right panels of Fig.\,\ref{fig combine}, the rates of period change for our SHeB sdB models are compared to the relative period changes measured by \cite{Pietrukowicz} using data from OGLE-III and OGLE-IV: $\dot{P}/P=\frac{\Delta P}{\Delta t}\frac{1}{P_{IV}}=\frac{P_{IV}-P_{III}}{t_{IV}-t_{III}}\frac{1}{P_{IV}}$. 

%
%
%

\subsection{Results}

As shown in Figure \ref{fig combine}, SHeB stars have typical rates of pulsation period change, $r$, on the order of $10^{-7}\rm yr^{-1}$, and occasionally on the order of $10^{-6}\rm yr^{-1}$. 
This is comparable to that of observed in confirmed and candidate BLAPs and high-gravtiy BLAPs.
Meanwhile, the values of $r$ may change from positive to negative (and vice versa), due to the fact that $P$ is proportional to the mean density and that $P$ increases or decreases when the star expands or shrinks. For example, for the models with $M_\mathrm{c}=0.45$ and $0.5M_\odot$, $r$ evolves from positive to negative when the stars evolve from expanding to shrinking during their evolution towards the white dwarf sequence. 
We note that, in most cases, the $r$ value is on the order of $10^{-7}yr^{-1}$ and the transitions from positive to negative are marked by $\Box$ in $T_{eff}-\log{g}$ diagrams. The open triangles and open squares in the left panels denote the positions where the transition occurs. The evolution of $r$ may account for the fact that both positive and negative $r$ values have been observed in BLAPs 
and this can be tested by future monitoring of $r$. 
However, the transition is hard to observe, as the sample size of BLAPs is too small. \footnote{A typical SHeB sdB stays $\sim 2\times 10^6$~yr in the BLAP region (see Sect 3.3) and if we observe for ten years, the possibility of observing the transition moment is about $10/(2\times 10^6)$. If we observe for ten years each of the 40 BLAPs, the possibility of observing the transition at least one time is $1-(1-(10/(2\times 10^6)))^{40} \approx 0.02\%$  }
We do not seen such transitions for models with higher core mass $M_{\rm c}=0.75$ and $1.0 M_\odot$ since these SHeB stars evolve directly to the post-AGB location and will not shrink until the He-rich envelope becomes thin enough. 
In fact, SHeB stars with $M_{\rm c}=0.75$ and $1.0 M_\odot$ have pulsation periods larger than observations and cannot cover the locations of high-gravity BLAPs on the $P-r$ plane. 

It is worth noticing that 
the radius fluctuates for models with $M_c = 0.5 M_{\odot}$,  resulting in large rates of period change ($r\sim 10^{-6}$) (panels in the second row). 
The large $r$ could explain BLAPs with positive large values such as OGLE-BLAP-011 and OGLE-BLAP-013.
We will investigate this into more detail in a future work.
\citet{Lin} reported the properties of TMTS-BLAP-1/ZGP-BLAP-01, and it is consistent with SHeB model but disagreed with other models.
In this paper, we simply assume that BLAPs and high-gravity BLAPs have the same types of pulsation and we regard the $0.45$ and $0.5 M_\odot$ models as our high-confidence ones. 

Figure 1 shows that SHeB models pass through the $T_{\rm{eff}}-\log(g)$ region inhabited by BLAPs 
when the H-rich envelop has the following masses: $M_e=0.013M_\odot-0.031M_\odot$ for $M_{\rm c}=0.45M_\odot$,
 $M_e=0.007M_\odot-0.011M_\odot$ for $M_{\rm c}=0.5M_\odot$, 
 $M_e=0.005M_\odot-0.037M_\odot$ for $M_{\rm c}=0.75M_\odot$, and 
 $M_e=0.009M_\odot-0.081M_\odot$ for $M_{\rm c}=1.0M_\odot$.
For the high-gravity BLAPs, we obtain $M_e=0.001M_\odot-0.013M_\odot$ for $M_{\rm c}=0.45M_\odot$, and
 $M_e=0.001M_\odot-0.005M_\odot$ for $M_{\rm c}=0.5M_\odot$.

Since our models are able to explain many properties of the BLAPs, in the following  discussion, we explore their possible formation channels. An important aspect that will be addressed in future works is to analyze the pulsation stability properties of (some of) our BLAP models. 
We remark that the study of the excitation and stability of pulsations strongly relies on both microscopic and macroscopic physical processes of elements, such as element diffusion, radiative levitation, and opacity tables \citep[for example][]{By}. We plan to analyze pulsation stability across the stellar parameters identified with our models in a future investigation by using non-adiabatic theory and radial stellar pulsation theory to try reproducing light curves and radial velocities observed in BLAPs.

The computation of non-adiabatic pulsations can provide powerful insights into the stability of pulsation modes. For example, in their work on pre-WD models to explain BLAPs, \cite{Romero} found the need of super-solar metallicity to drive pulsations via the $\kappa$ mechanism.  The effect of metallicity on the stability of pulsation is particularly interesting, given the lack of BLAPs detected thus far in the metal-poor regime \citep{Pietrukowicz2018}. Since the fraction of close binaries increases at low metallicities \citep{Moe}, if our models are not pulsationally unstable at low metallicity, this could have implication for the viability of evolutionary scenarios relying on close period binaries (see the next section). 

\section{Formation channel(s)}

Given the success of our SHeB sdBs models in reproducing a number of BLAPs' properties, in this section we investigate their formation channel(s). \citet{Han2002,Han2003} developed a binary model for the formation of sdB stars. 
This model explained almost all properties of sdB stars, including single sdBs, short- and long-period sdBs\footnote{For the long-period sdBs see \citet{Chen2013,Vos2019,Vos2020}},
and thus has been widely accepted in the literature.  
Our study for BLAPs is based on this model.

Except for HD 133729 \citep{Pigulski}, no evidence for companion stars has been found in BLAPs so far \citep{Pietrukowicz,Kupfer,Ramsay,McWhirter2020}. This however 
does not necessarily imply that BLAPs are single stars. For example, they could be in long-period binaries which have escaped detection or, otherwise, in short-period binaries, where very bright companions dominate the light budget and dilute the pulsation amplitude (similar to HD 133729, see \citealt{Pigulski}). It might also be possible that BLAPs themselves have very faint companions. Radial velocity monitoring might be used to detect binaries, but so far only few BLAPs have spectroscopic  observations.

In Han's model, hot subdwarfs can be produced through three channels, namely, CEE, stable RLOF, or a merger of two He WDs \citep{Han2002,Han2003}.
For the CEE channel, 
the donor loses most of the H-rich envelope mass close to the tip of RGB during common envelope ejection and the remnant evolves to short-period sdBs. This channel is inconsistent with the -albeit limited- observations that no BLAP shows sign of being in a short-period binary, except for the case of the companion of HD 133729. 
Since the CEE process is complicated and is not yet well understood \citep[see][]{Ivanova}, 
it is hard to know the envelope mass of sdBs in this way from binary calculation. However, the locations of observed short-period sdBs in the $T_{\rm eff}-{\rm log} g$ diagram suggest that 
the H-rich envelopes of sdBs from the CEE channel are massive enough to account for BLAPs \citep{Xiong}.
If we only consider the envelope mass of sdBs, we cannot exclude the possibility that the CEE produces BLAPs\footnote{If sdBs from the CEE channel cannot evolve into the BLAPs, 
the abundances in the envelope could be the cause since element abundances are crucial 
for driving the pulsations. There is evidence showing that short-orbital period sdBs
resulted from the CEE channel have surface abundances that are significantly different from that of the RLOF \citep{Geier2022}.}. Therefore, in Section 3.3, we give the expected number of BLAPs also from this channel. 

As we have already pointed out, HD 133729 is the only object for which a BLAP has been identified in a binary system and given its parameters it was likely produced via  stable RLOF rather than CEE. The (present-day) primary in HD 133729, has a mass of $2.85 \pm 0.25 M_\odot$, which means that the donor (which we speculate to have evolved into the BLAP) probably had a mass around $\sim 3.0 M_\odot$ \citep{McWhirter} and non-degenerate core before the RLOF took place. Such a star may become an sdB star (and then a BLAP) if mass transfer starts during the Hertzsprung gap. Based on the work of \citet{Han2003}, sdB binaries produced by non-degenerate progenitors via RLOF during a Hertzsprung gap have orbital periods in the range between 10 hours and dozens of days, consistent with the observations of HD 133729.


For the merger channel, two He WDs could merge due to gravitational wave radiation
and produce a single hot subdwarf star, with its mass in the range of $ 0.3-0.8 M_\odot$ \citep{Han2002,Han2003,Zhang2009,Zhang2017}. 
During the merger of double He WDs, the less massive He WD is disrupted and accreted onto the more massive one.
Helium is ignited through He flashes, and the merged object becomes an sdB star when He burns stably in the center \citep{Zhang2012}. In this case, however, little hydrogen survives the series of He flashes that is 
the maximum hydrogen mass is only 0.002$M_\odot$ in the study of \citet[][see also \citealt{Zhang2012}]{Philip2016}. This value agrees with the $M_\mathrm{e}$ determined for high-gravity BLAPs but it is far less than that required for the other BLAPs ($>0.005M_\odot$, as discussed in Section 2). Hence, if BLAPs are single sdBs, 
it is unlikely that they form through the merger of two He WDs\footnote{However, we cannot exclude the possibility that BLAPs are single stars according to this. For example,  \citet{Meng} proposed that BLAPs are possibly the surviving companions of type Ia supernovae with masses of $ 0.7-1.0 M_\odot$.}.

For the RLOF channel, if the donors have low initial masses ($\lesssim 2M_\odot$) and degenerate He cores after central H burning, 
the produced hot subdwarfs have masses around $0.5M_\odot$ \citep{Han2002,Han2003} and orbital periods around 1400 days \citep{Chen2013}. 
If the donors have masses with non-degenerate He cores\footnote{Although it is typically quoted that non-degenerate cores form beyond $2.0M_\odot$, with the models used in this work this happens for $>1.99M_\odot$.}, 
the resulting hot subdwarfs have a wide mass range, namely, from $0.3-0.8M_\odot$,
and the orbital periods are relatively short, namely,
from several days to more than one hundred days,
depending on the initial mass ratio, initial orbital period, and the assumptions for mass and angular momentum loss \citep{Han2000,Chen2002,Chen2003}. 
In the following, we focus on the RLOF channel with donors ranging from evolved to degenerate He cores and we create long orbital period binaries.


\subsection{Binary evolution calculations}

Using MESA, we investigate the binary evolution for several binaries consisting of a giant star (the progenitor of sdBs) and an MS or WD companion.
The study is based on Population I stars and the basic physics inputs are the same as introduced in Sect. 2. 
We only evolve donors and consider companions as point sources. In our calculation, the mass transfer rate is calculated by the scheme of \cite{Kolb},  
and the mass transfer process is completely non-conservative, 
that is, all mass lost from the donor onto the MS star is not actually accreted by the MS star, but is lost from the MS and, hence, from the system. 
The angular momentum lost from the system is then given by the mass loss times the specific angular momentum (i.e., angular moment per unit mass) of the MS star. No wind loss has been included in the whole evolutionary process. We discuss this point further in Sect. 3.4.

We go on to explore eight values for the initial mass of the giant (the donor: $M_{\rm 1i}= 0.79, 0.89, 1.0, 1.12, 1.26, 1.4, 1.58$, and$ 1.78M_\odot$) and three values for initial giant and MS mass-ratio $q_{\rm i}=1.1, 1.25$ and 1.5.
The upper limit of $q_{\rm i}=1.5$ here is consistent with the critical mass ratio 
for dynamically stable mass transfer when the donor is a giant \citep{Han2002,Chen2008}.
For each ($M_{\rm 1i}$, $q_{\rm i}$), we increase the initial orbital separation, $A_{\rm i}$, in steps of equal $\Delta A_{\rm i}$, from the minimum separation to produce sdB stars, to the point where the donor cannot fill its Roche lobe on the red giant branch.

\begin{figure}
\includegraphics[width=8cm]{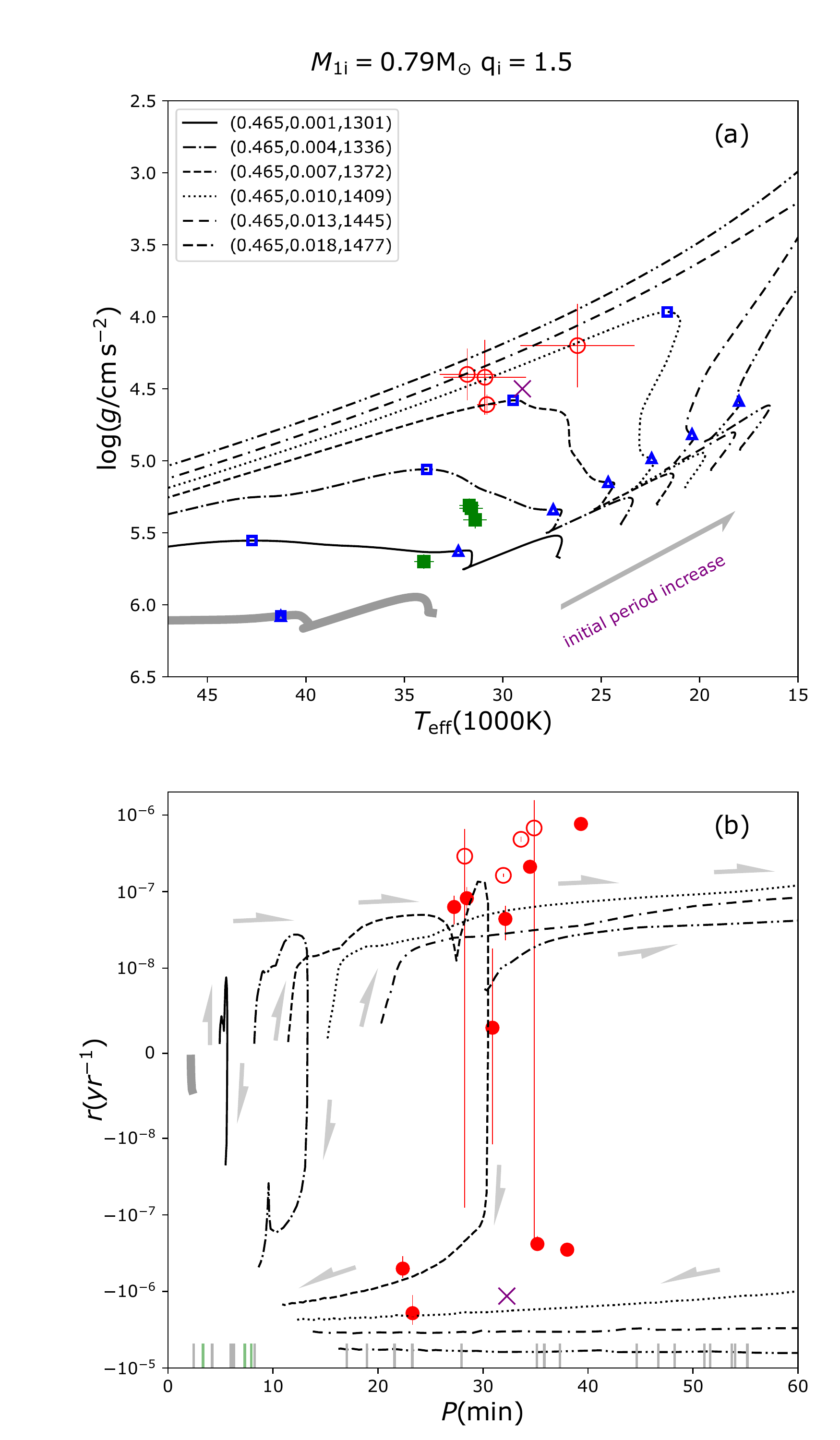}
\caption{Evolution of sdBs produced via binary evolution for ($M_{\rm 1i}, q_{\rm i}$)=(0.79$M_\odot$,1.5). The thick gray line identify overlapping evolutionary tracks (there are five tracks overlapping in this figure, see text for details).  
Panel (a): Evolutionary tracks in the $T_{\rm eff}-{\rm log} g$ plane. Each linestyle corresponds to a different set of $(M_c,M_e,P_{orb})$ at the start of stable CHeB (see legend).
Panel (b):  Rate of period change, $r,$ versus pulsation period, $P,$ for the models in panel (a). 
The final orbital periods for these models are $\sim 1400$ days.
Observations are the same as in Fig.\,\ref{fig combine}.}
\label{fig3}%
\end{figure}

\subsection{Outcomes from binary evolution}

From our binary evolution calculations, we obtain in total 24 sets 
of long-period SHeB sdBs. 
For each ($M_{\rm 1i}, q_{\rm i}$), we obtain some SHeB sdBs passing through the position of BLAPs and that of high-gravity BLAPs. 
We show an example in Fig.\,\ref{fig3},
with initial parameters ($M_{\rm 1i}, q_{\rm i}$)=($0.79M_\odot$,1.5): the upper panel displays evolutionary tracks in the $T_{\rm eff}-{\rm log} g$ plane, while the lower one shows those in the $P-r$ diagram.
A total of 11 evolutionary tracks (including 5 tracks overlapping in the thick gray line) are shown in Fig.\,\ref{fig3}, with initial orbital periods ($\sim$ orbital periods before RLOF) 
equally spaced in terms of log.
When the initial orbital periods are short, 
the produced sdBs nearly have no H-rich envelope due to delayed He flashes \citep{Xiong} 
and the evolutionary tracks are overlapped, as shown with thick grey line in the upper panel.

As shown in the upper of Fig.\,\ref{fig3}, several SHeB sdBs with relatively high envelope mass match the location of BLAPs, while some with relatively low envelope mass match the location of high-gravity BLAPs, similarly to what has already been shown in Fig.\,\ref{fig combine}.
The pulsation period and its rate of change match the observations in terms of the right order of magnitude.  
All models have orbital period of $\sim 1400$ days and the companion is a $0.53M_\odot$ main sequence star due to no accretion in the RLOF process (mass transfer is completely nonconservative).

The results for the other explored sets of ($M_{\rm 1i}, q_{\rm i}$) are qualitatively similar to those shown here, (presented in Figs. A1-A23 in the appendix. All sdBs we produce have very similar evolutionary tracks in both the $T_{\rm eff}-{\rm log} g$ and $P-r$ diagrams. They also have orbital periods greater than 1000 days, with companions in the mass range  of $0.53 - 1.62 M_{\odot}$.

We remark that it is somewhat difficult for our models to explain the relatively high positive $r$ values observed in some BLAPs; however, we can clearly see in Fig.\,\ref{fig combine} that some models may reproduce it. This happens more easily for high He-core masses $M_{\rm c}=0.75M_\odot$ and $1.00M_\odot$, and even for $M_{\rm c}=0.5M_\odot$ when fine-tuning envelope masses. This seems to suggest that RLOFs with non-degenerate donors might also be a viable channel for BLAP formation and they are worthy of more investigations in a future work.

\subsection{Number of BLAPs in the Galaxy}
 According to our calculation, about 25 to 50 percent (marked with $r_{\rm 1}$) of sdBs would evolve to become BLAPs during their lives
\footnote{We have computed a total of 304 evolution tracks and between $\sim$ 80 and $\sim$ 160 pass through the location of BLAPs. This range is due to the uncertainty on the location of BLAPs in the $T_{\rm eff}-{\rm log} g$ diagram. For example, the first 14 BLAPs discovered by \cite{Pietrukowicz} cover a much smaller region than that reported in the most recent works.}. 
Considering that the typical lifetime of sdBs is $t_{\rm sdB} \sim 10^8 \rm{yrs}$ 
and the lifetime of SHeB sdBs 
is $t_{\rm shell} \sim 5 \times 10^7 \rm{yrs}$,
we can simply estimate the number of SHeB 
sdBs produced in this way as $N_{\rm shell} = r_{1}*N_{\rm sdB}*t_{\rm shell}/t_{\rm sdB}$, where $N_{\rm sdB}$ is the number of sdBs produced from the RLOF channel in the Galaxy. 

From the study of \citet{Han2003}, 
$N_{\rm sdB}=4.36\times10^{6}$. 
We therefore have 
$N_{\rm shell} \sim 0.5 - 1 \times10^{6}$.
The number of BLAPs depends on the parameter space defined for BLAPs. 
A typical SHeB sdB stays $\sim 2\times10^{6}$ yrs in the BLAP region defined by \citet{Pietrukowicz}, only about $1/20$ (marked with $r_{\rm 2}$) of the SHeB sdB time.
Also, not all stars in the instability region show pulsations. 
For example, ten percent of subdwarfs in the rapid variability region \citep[sdBVr,][]{Ostensen} and 75 percent in the slow variability region \citep[sdBVs,][]{Green} show pulsations. We hence assume the fraction of stars showing pulsation to be in the range $r_{\rm 3}=0.1-1$. 
The number of BLAPs produced through the RLOF channel therefore can be estimated as $N_{\rm BLAPs}=r_{\rm 2}*r_{\rm 3}*N_{\rm shell}$, ranging from $\sim 2500$ to $\sim 50000$. 

Observationally, we can estimate the number of BLAPs from existing data. For example, \citet{McWhirter} discovered 22 BLAPs candidates (six of which are high-gravity ones) by cross-matching Gaia DR2 and ZTF DR3 using a sample of over 162 million sources satisfying a number of photometric and astrometric cuts. Excluding for simplicity the impact of these cuts on sample selection, a plain scaling to the number of stars in the Milky Way (which is in the range 100-400 billion stars), gives an estimated number of BLAPs candidates between $10^4$ and $10^5$. If half of them are BLAPs, their number should be of a few tens of thousands, which finely matches our estimates obtained from 
SHeB sdBs formed through RLOF channel, although uncertainties are still admittedly large. 
Our estimates are also consistent with the range 280 to 28000 derived by \cite{Meng} from the OGLE survey under the two extreme assumptions that all BLAPs had been discovered, or that 99 percent of them had been missed, respectively. 

If we consider that the CEE channel may also contribute to the formation of BLAPs, and assume sdBs from CEE have an envelope mass as similar to that from RLOF, from the study of 
\cite{Han2003}, we have $N_{\rm sdB}=6.18\times10^{6}$ for both the CEE+RLOF channels. Keeping the other assumptions unchanged, the estimated $N_{\rm BLAPs}$ ranges from $\sim 3500$ to $\sim 70000$, still consistent with the observations. 
In further work, a detailed stellar population synthesis \citep[like][and Alexey et al. in prepared]{Byrne2021} may give a more precise number and the distribution of properties of BLAPs.

\subsection{Putative BLAPs companions}

In our calculations, all SHeB sdBs likely to be observed as BLAPs 
have relatively long orbital periods, $P \sim 1400$ days. 
The companions are main-sequence stars with lower mass limit, 
since the mass transfer is set to be completely non-conservative and the companion will not accreate any material during the RLOF.
In this section, we explore and quantify the impact of such companions on photometry and radial velocities. 

\begin{figure}
\includegraphics[scale=0.45]{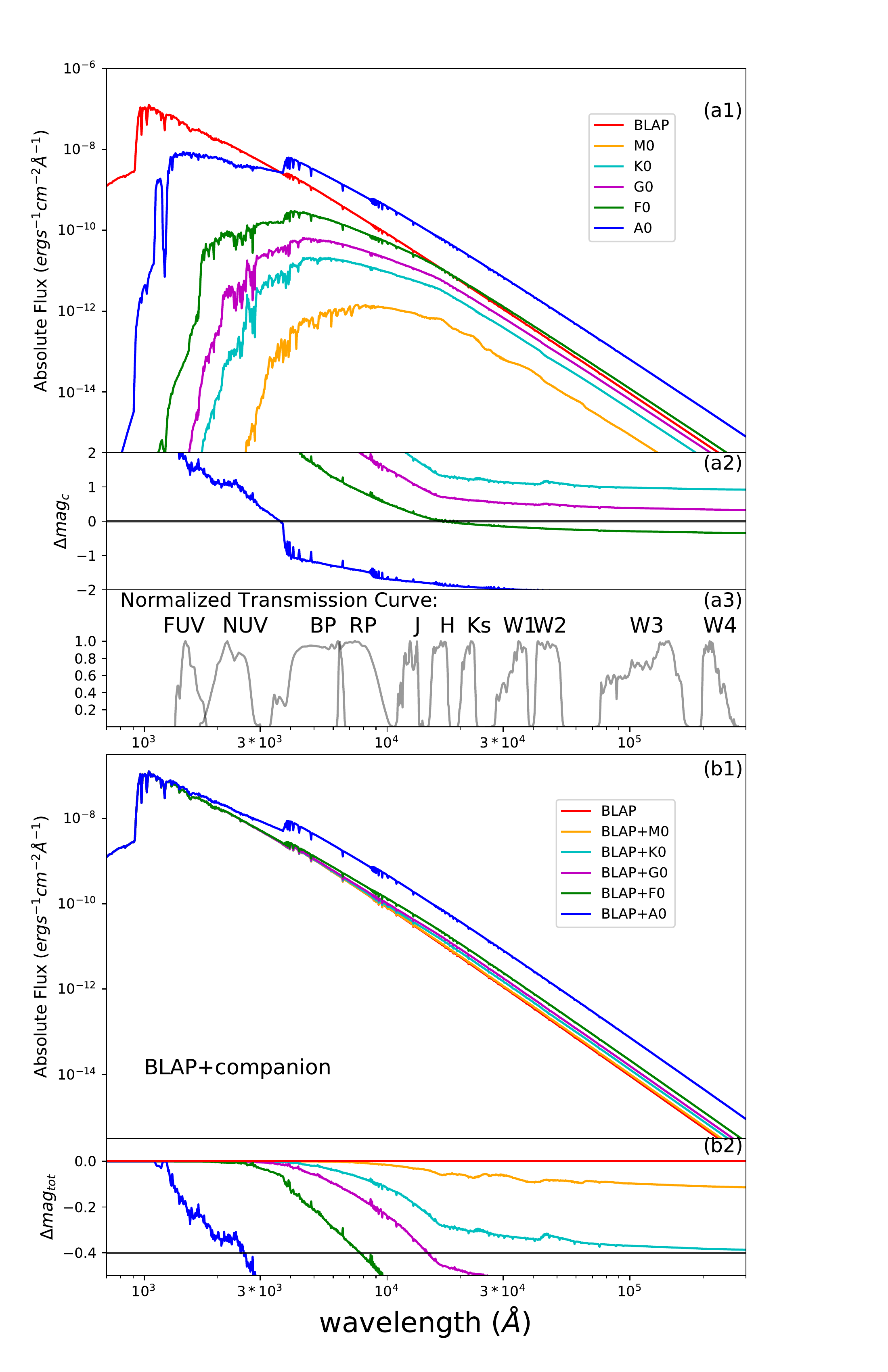}
\caption{The expected differences in SED between a BLAP have a companion or not.Panel {\it a1} ({\it b1}): Individual (combined) spectral energy distributions for the BLAP and main sequence companions listed in Table \ref{tab:param}. Panel {\it a2} and {\it b2}: Monochromatic magnitude differences. See text for details. 
The filters transmission curves are $FUV$ and $NUV$ from GALEX, $BP$ and $RP$ from Gaia, $J$, $H,$ and $K_s$ from 2MASS and $W1$, $W2$, $W3$, and $W4$ from WISE.
\label{fig:sed}}%
\end{figure}

In Fig.\,\ref{fig:sed}, we show the spectral energy distribution (SED) of a star with BLAP-like stellar parameters, in comparison to typical A- to M-type main sequence stars. 
We use synthetic fluxes at solar metallicity from the grid of \cite{ck03}, rescaled by the square of the adopted stellar radius and interpolated at the appropriate $T_{\rm{eff}}$ and $\log{g}$ (see Table \ref{tab:param} for the adopted values). All stars are assumed to be at the same distance. 
For comparison, we also plot a number of filter transmission curves from the ultraviolet to the mid-infrared. 
From panel (a1) we can clearly see that 
for most spectral types the flux of the BLAP is dominant over the main-sequence companion. 
Sufficiently hot main sequence stars, however, can become dominant in the optical (A-type) or in the infrared (F-type), 
depending on the adopted radii. 
Panel (a2) shows similar comparison, but this time for 
the monochromatic magnitude difference between the flux of the main-sequence companion and that of the BLAP. 

Observationally, however, this hypothetical binary systems must have been unresolved so far. This means that only the total flux can be observed. To account for this, in panel (b1) we compare the flux of a single BLAP (red) against that of a BLAP with a companion. The slope of the SED in the ultraviolet is largely determined 
by the $T_{\rm eff}$ of the BLAP, but it is altered by the flux of a companion at increasingly longer wavelengths. This excess amounts to $\sim 0.1$ mag for an M companion in the mid-infrared and it is considerably larger for earlier spectral types. However, it must be kept in mind that the effective temperature and the radius of a BLAP might vary quite considerably while pulsating, thus  complicating the detectability of potential companions.  



\begin{table}
\caption{Physical parameters adopted for the SED of main sequence stars orbiting a BLAP. For the latter, we adopted the values from one of our binary models. For main sequence companions, we used literature parameters for an archetypal spectral types \citep{zombeck07}.}
 \label{tab:param}
\centering
\begin{tabular}{cccccl}     
\hline\hline
 Spec. type & $T_{\rm{eff}}({\rm K})$& $\log{g}$  &$R(R_{\odot})$&$M(M_{\odot})$& archetypal\\
 \hline
   BLAP & 27718 & 4.51 & 0.62 & 0.469 & \\
   A0   & 10800 & 4.15 & 2.50 & 3.20  & $\alpha$ CrB A\\
   F0   & 7240  & 4.44 & 1.30 & 1.70  & $\gamma$ Vir \\
   G0   & 5920  & 4.44 & 1.05 & 1.10  & $\beta$ Com \\
   K0   & 5240  & 4.47 & 0.85 & 0.78  & 70 Oph A \\
   M0   & 3800  & 4.80 & 0.51 & 0.60  & Lacaille~8760 \\
 \hline\hline
\end{tabular}
\end{table}

Now let us considered the possibility of looking for binarity through radial velocity variations. From our models, a BLAP with mass 0.458 $M_{\odot}$ orbiting a companion of 1.62 $M_{\odot}$ on a period of 1366 days will have an orbital speed of 19 km/s, assuming circular orbit. A model with a significantly less massive primary on similar period (0.465, 0.72, 1372) has an orbital speed of 11 km/s. These values will decrease with the inclination angle at which the system is observed. If BLAPs are instead composed of binaries with short periods, the radial velocity signal will be significantly higher. It is thus reasonable to aim for a radial velocity precision on the order of one km/s to test whether BLAPs have companions.  

\begin{figure}
\includegraphics[width=9cm]{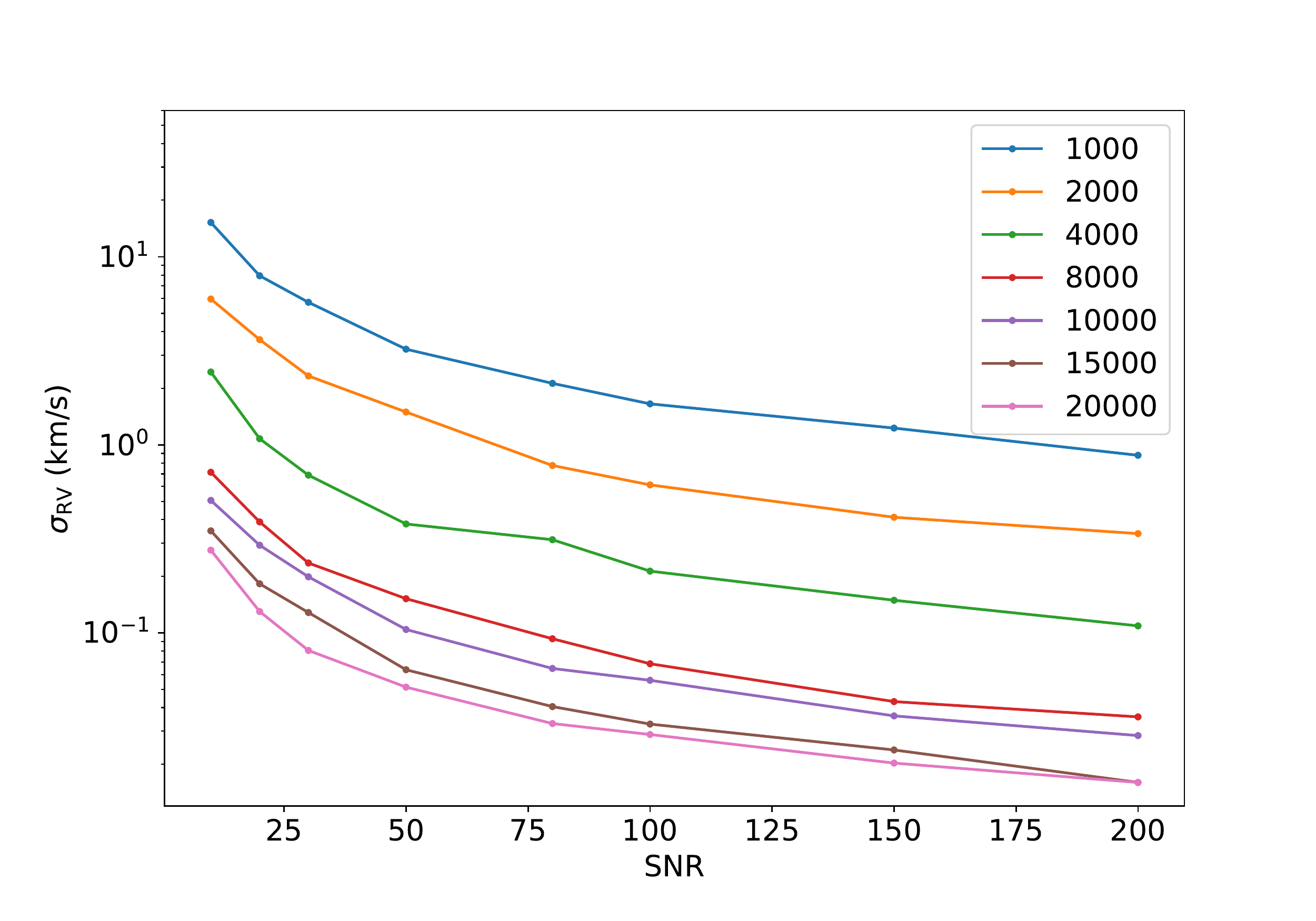}
\caption{Errors (standard deviations) of the radial velocities measured by using the method of cross-correlation function for different spectrum resolution (R) and S/N. Colors indicate the resolution.
\label{fig:rvsigma}}%
\end{figure}

To explore this possibility in more detail, we simulated the radial velocity precision that can be achieved with spectra of various resolution (R) and signal-to-noise ratio (S/N). For the sake of simplicity, we fixed the parameters of the BLAP at $T_{\rm eff}=30,000$ K, $\log{g}=4.5$ dex, and solar metallicity using a TLUSTY synthetic spectrum \citep{Lanz2003}. We generated 100 spectra in the wavelength range $3800 - 7400\ $\AA$\ $for several values of resolution and signal-to-noise, and then estimated the radial velocity precision by computing the standard deviations of the radial velocities that are measured via cross-correlation function \citep{Tonry} As shown in Fig.\,\ref{fig:rvsigma}, radial velocity errors decrease with increasing spectral resolution and signal-to-noise, achieving a km/s precision when R > 4000 and S/N > 25.
It should, however, be pointed out that our results are idealized as they do not include uncertainties arising from instrumental effects, nor the contribution from the pulsations of a BLAP. These would dominate the radial velocity error when integrating over a pulsation period, although stable radial velocity monitoring over a long period of time would still be able to detect the modulation due to a companion over the short period pulsations.

\subsection{Discussion}
Our study assumes that the mass transfer processes in this part of parameter space are completely nonconservative, namely, that all the mass lost from the giant is lost from the system. 
If the mass transfer is somehow conservative, the results of RLOF will change.
Also, different assumptions for mass and angular momentum loss 
will result in different parameter spaces for producing sdB stars 
that is, the range of $A_{\rm i}$ for a given $M_{\rm 1i}$ and $q_{\rm i}$.
However, as donors in our study have degenerate cores, the results satisfy a unique sdB mass - orbital period relation \citep{Chen2013}. This means that the orbital period will not change if the sdB mass stays constant, and only the mass of the companion changes. 
Therefore, the sdB binaries formed through RLOF channel have the same characteristics of our results (i.e., core mass, envelope mass, and orbital period), although the companion would become more massive than that given in this study and may become a blue straggler \citep[see][]{Chen2008}. 
And a more massive companion would be easier to be detected via SED and the BLAP would have a higher radial velocity (as described in Sect. 3.4).

We have not considered the case where the companion is a WD in our model.
We explain this as follows. 
To ensure the mass transfer process is dynamically stable, 
the pre-mass-transfer mass of the giant is very restricted, that is, the mass ratio of the giant to the WD is expected to be below a value of $\sim 1.1-1.3$ \citep[see table 3 of][]{Han2002}. 
This generally requires very massive WD companions. 
Since massive WDs are very rare, they are unlikely to be the companions of BLAPs. 
In fact, \cite{Han2003} have not obtained any sdB stars with a WD companion in their model.
However, based on an adiabatic mass loss model, \cite{Ge2013,Ge2015} obtained significantly larger critical mass ratios for the stability of mass transfer when the donor is on the giant branch, indicating that the WD mass could be lower than believed for stable mass transfer.
We estimated the impact of the critical mass ratios and obtained a few sdB+WD binaries from the RLOF channel indeed based on the criterion of the Ge et al.(2020). 
But the number is very small and the WD companions to BLAPs should consequently be very rare. 
For the case of neutron star (NS) or black hole companions, 
the mass transfer is more likely to be stable due to massive companions. 
According to the study of \citet{WuYou}, one percent of sdBs have NS companions with long periods ($\sim 1000$ days), we thus may expect $\sim 20-500$ BLAPs with NS companions -- 
in a similar way to what is explained in Section 3.3.

The CEE channel will generate sdBs in short period binaries. However, it is still an open question how much H-rich envelope will be left after the CEE \citep{Xiong}, due to the huge uncertainties in common envelope evolution \citep[see][]{Ivanova}. 
Presently we cannot rule out the possibility that some sdBs from CEE have relatively thick H-rich envelopes and show characteristics similar to BLAPs. 
Their companions could be discovered from radial-velocity variations (see footnote 8). 
HD 133729 proves that BLAPs in short-period binary do exist, implying that the CEE channel might be a viable formation scenario. Because the primary is a B-type main sequence with $\sim 2.5 M_\odot$, the progenitor of this BLAP would be an intermediate mass star, which is outside of our grid. According to \cite{Han2003}, HD 133729 also could be produced by the RLOF channel, in which the RLOF occurs on Hertzsprung gap. We will investigate this possibility in future work.

\section{Conclusions}

In this paper, we explore whether SHeB sdBs can reproduce the properties of BLAPs and we discuss their formation channels, based on binary evolution. 
Our study has shown that some SHeB sdBs aptly reproduce the properties of BLAPs, namely, effective temperature, surface gravity (or luminosity), and the rate of period change. Given that the purpose of this paper is to explore a new evolutionary pathway to BLAPs and narrow down its parameter space, we made use of simple scaling relations to infer the pulsation period and rate of period change for our models.  
The best-fitting sdB models are in the mass range $0.45-0.5M_\odot$, which can explain both BLAPs and high-gravity BLAPs. On the contrary, models with higher masses ($0.75-1.0M_\odot$) are not able to reproduce the properties of high-gravity BLAPs. A prediction of our modeling is that positive $r$ values may change into negative ones. 

We  performed a series of binary evolution calculations to investigate the formation of BLAPs through the RLOF channel. 
From our grid of 304 models, we have found that 142 SHeB sdBs could be good BLAP candidates. All these BLAP candidates have long orbital periods, that is, $\gtrsim 1400$ days. Because of these periods and the faintness of known BLAPs, the detection of companions would require dedicated radial velocity monitoring. Main sequence companions might also be detected through their flux excess when carrying out the SED fitting.

We also briefly discuss how CEE and the double He WDs merger scenarios might produce sdBs with characteristics similar to BLAPs. The CEE channel preferentially leads to short-period binaries, which would be easier to detect than long-period ones through radial velocity monitoring. Since the fraction of close binaries increases at low metallicity \citep[][]{Moe}, pulsation stability in the metal poor regime might give important constraints on the suitability of such a scenario. 
The merger of double He WDs leads instead to the formation of single sdBs with very thin hydrogen envelopes -- far less than required for BLAPs, but consistent with high-gravity BLAPs. We thus conclude that dedicated radial velocity monitoring of BLAPs could help to unveil the origin of these mysterious objects. 

\section*{Acknowledgements}

The authors gratefully acknowledge insightful comments by P. Pietrukowicz, which helped to improve the paper. We thank the anonymous referee for their comments which have helped to improve the paper. This work is partially supported by the Natural Science Foundation of China (Grant no. 11733008, 12125303, 12090040/3),the China Manned Space Project of No. CMS-CSST-2021-A10. J. Vos acknowledges support from the Grant Agency of the Czech Republic (GA\v{C}R 22-34467S). The Astronomical Institute Ond\v{r}ejov is supported by the project RVO:67985815. S. Justham acknowledges funding from the Netherlands Organisation for Scientific Research (NWO), as part of the Vidi research program BinWaves (project number 639.042.728, PI: de Mink).
T. Wu thanks the supports from the B-type Strategic Priority Program of the Chinese Academy of Sciences (Grant No. XDB41000000), from the National Key R\&D Program of China  ( Grant No. 2021YFA1600402), from the NSFC of China (Grant Nos.  11873084, 12133011, and 12273104), from Youth Innovation Promotion Association of Chinese Academy of Sciences, and from Ten Thousand Talents Program of Yunnan for Top-notch Young Talents. 

\bibliographystyle{aa}

\bibliography{heran.bib}


\begin{appendix} 

\section{$T_{\rm eff}-{\rm log} g$ and $P-r$ diagram of SHeB SdBs produced by RLOF channel}
The evolutionary tracks on $T_{\rm eff}-{\rm log} g$ and $P-r$ diagrams from binary evolution (sea Sect. 3.2). In Sect. 3.2, we only display the results for one ($M_{\rm 1i}, q_{\rm i}$), here, we show other results.

\begin{figure}
\includegraphics[width=8cm]{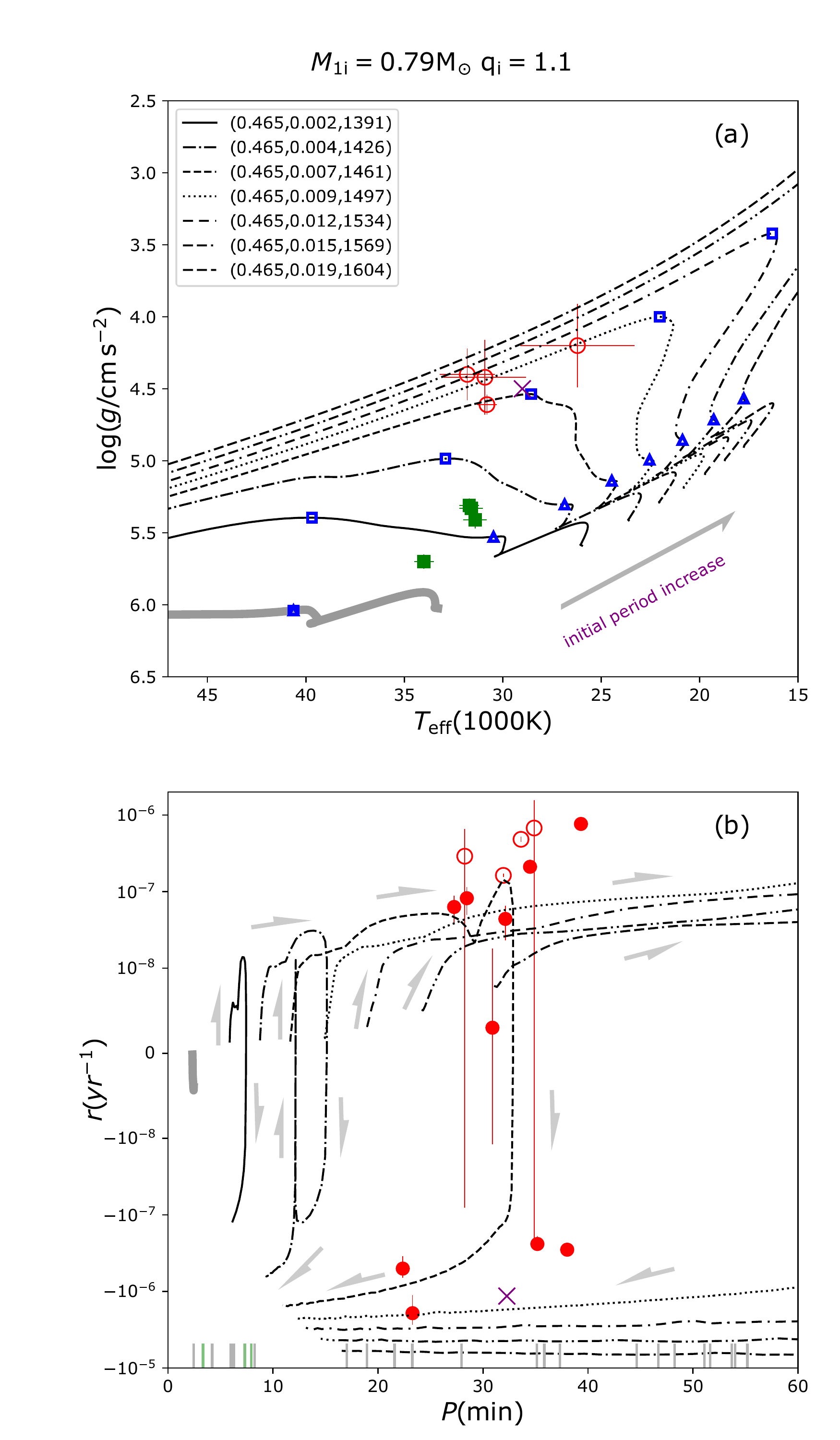}
\caption{Evolution of sdBs produced by binaries, which is same as in Fig.\,\ref{fig combine}, but for ($M_{\rm 1i}, q_{\rm i}$)=(0.79$M_\odot$,1.1).
\label{fig A1}}%
\end{figure}

\begin{figure}
\includegraphics[width=8cm]{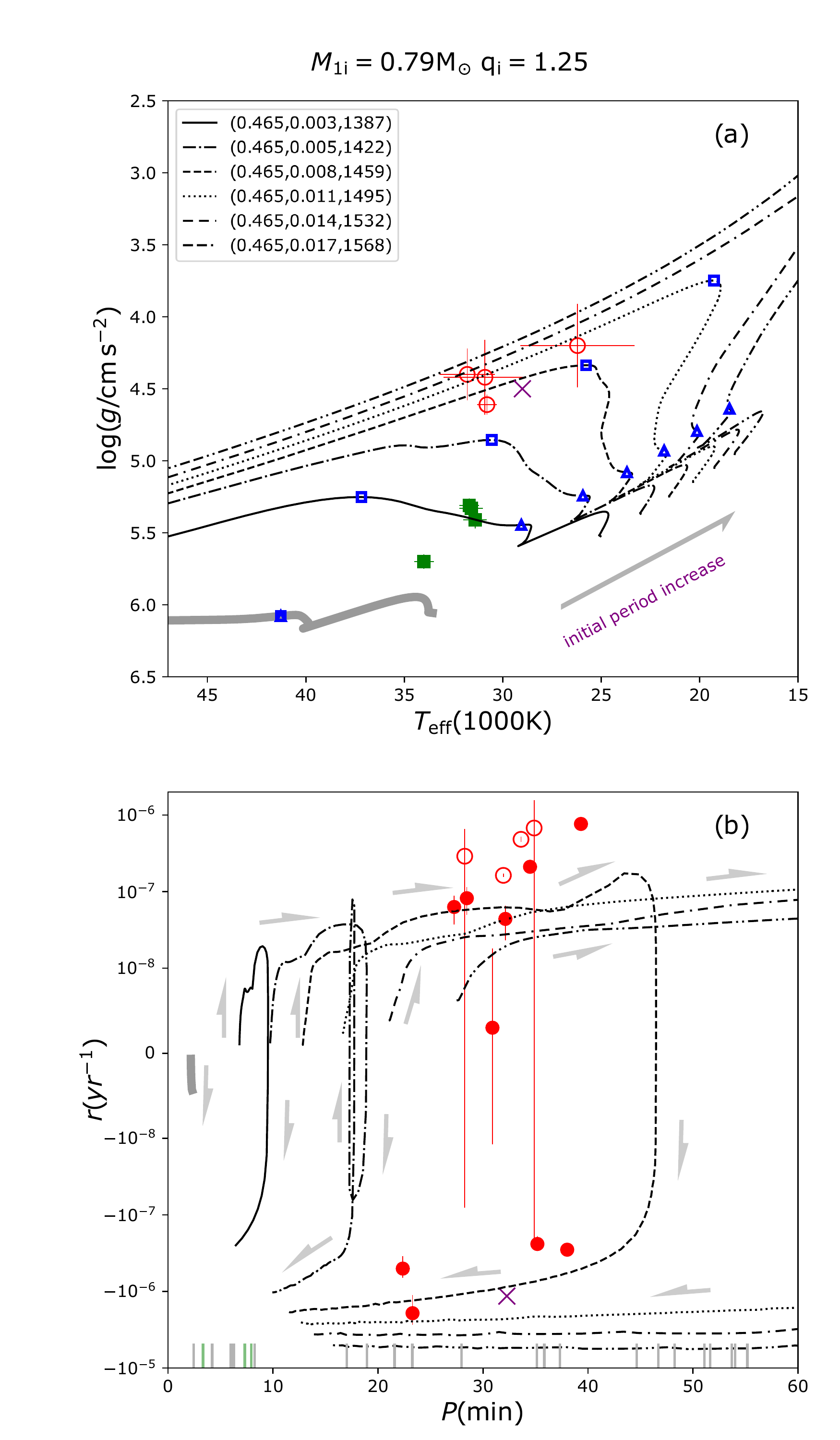}
\caption{Evolution of sdBs produced by binaries, which is same as in Fig.\,\ref{fig combine}, but for ($M_{\rm 1i}, q_{\rm i}$)=(0.79$M_\odot$,1.25).
\label{fig A2}}%
\end{figure}

\begin{figure}
\includegraphics[width=8cm]{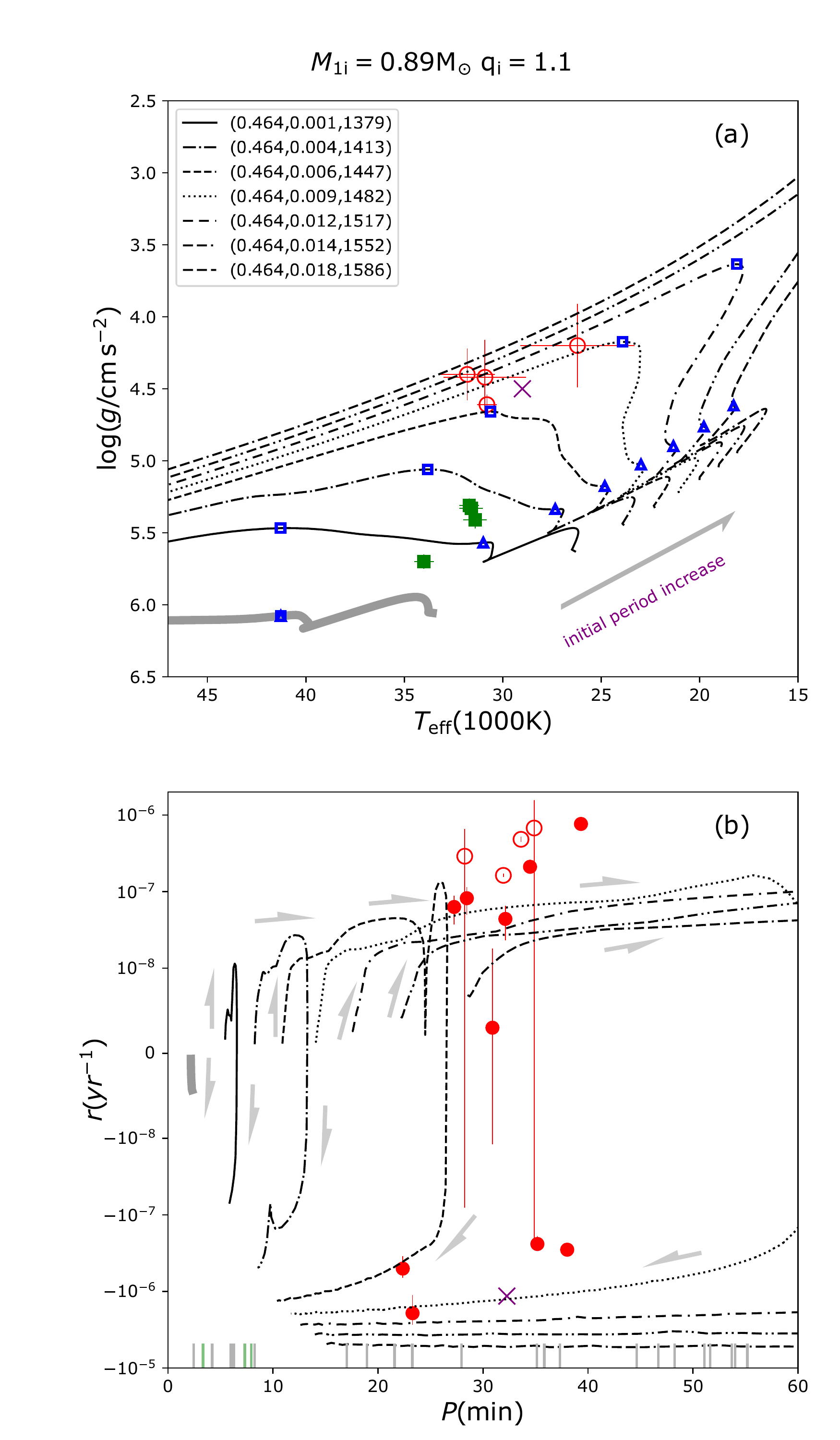}
\caption{The evolution of sdBs produced by binaries, which is same as in Fig.\,\ref{fig combine}, but for ($M_{\rm 1i}, q_{\rm i}$)=(0.89$M_\odot$,1.1).
\label{fig A3}}%
\end{figure}

\begin{figure}
\includegraphics[width=8cm]{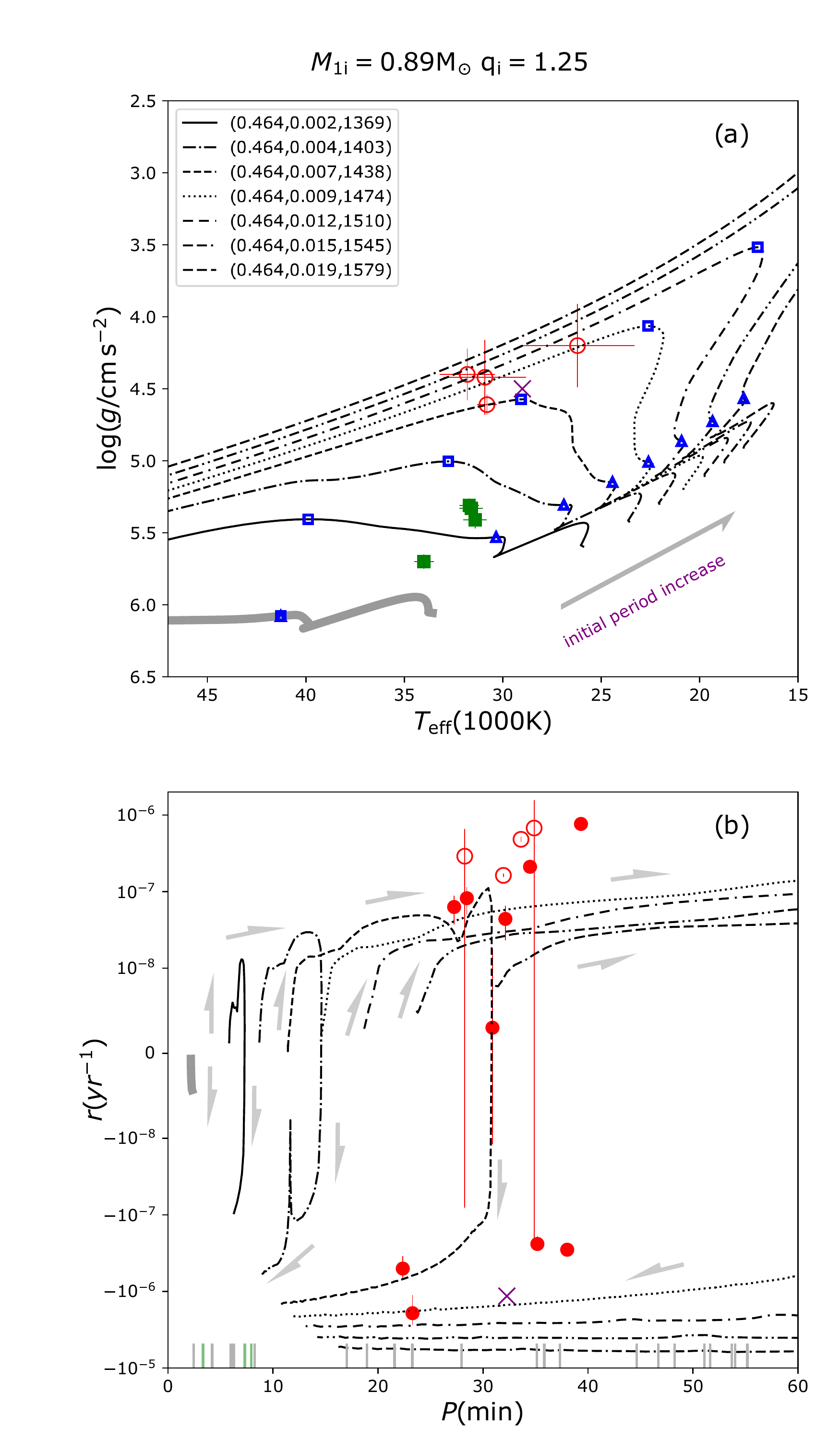}
\caption{The evolution of sdBs produced by binaries, which is same as in Fig.\,\ref{fig combine}, but for ($M_{\rm 1i}, q_{\rm i}$)=(0.89$M_\odot$,1.25).
\label{fig A4}}%
\end{figure}

\begin{figure}
\includegraphics[width=8cm]{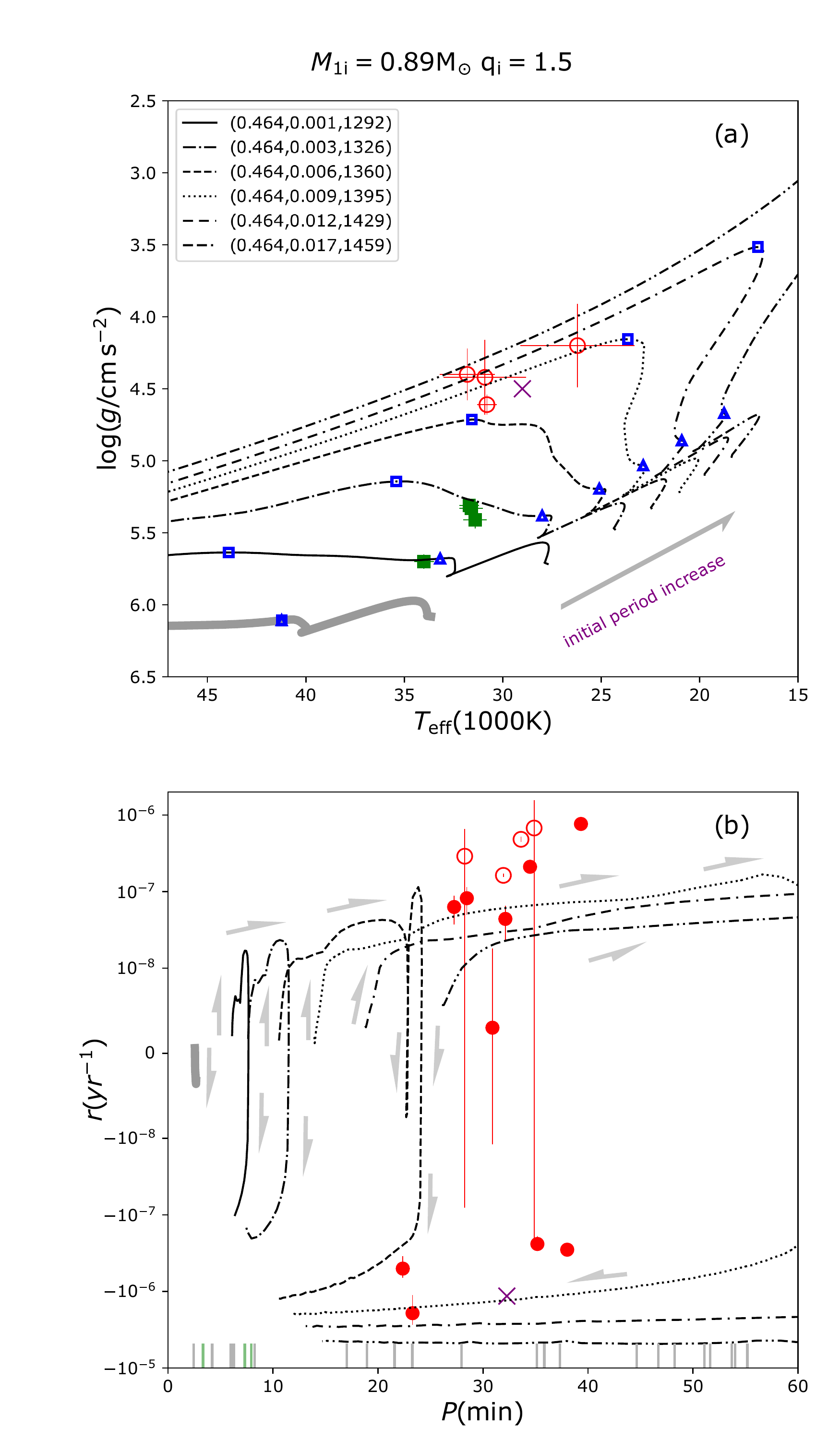}
\caption{Evolution of sdBs produced by binaries, which is same as in Fig.\,\ref{fig combine}, but for ($M_{\rm 1i}, q_{\rm i}$)=(0.89$M_\odot$,1.5).
\label{fig A5}}%
\end{figure}

\begin{figure}
\includegraphics[width=8cm]{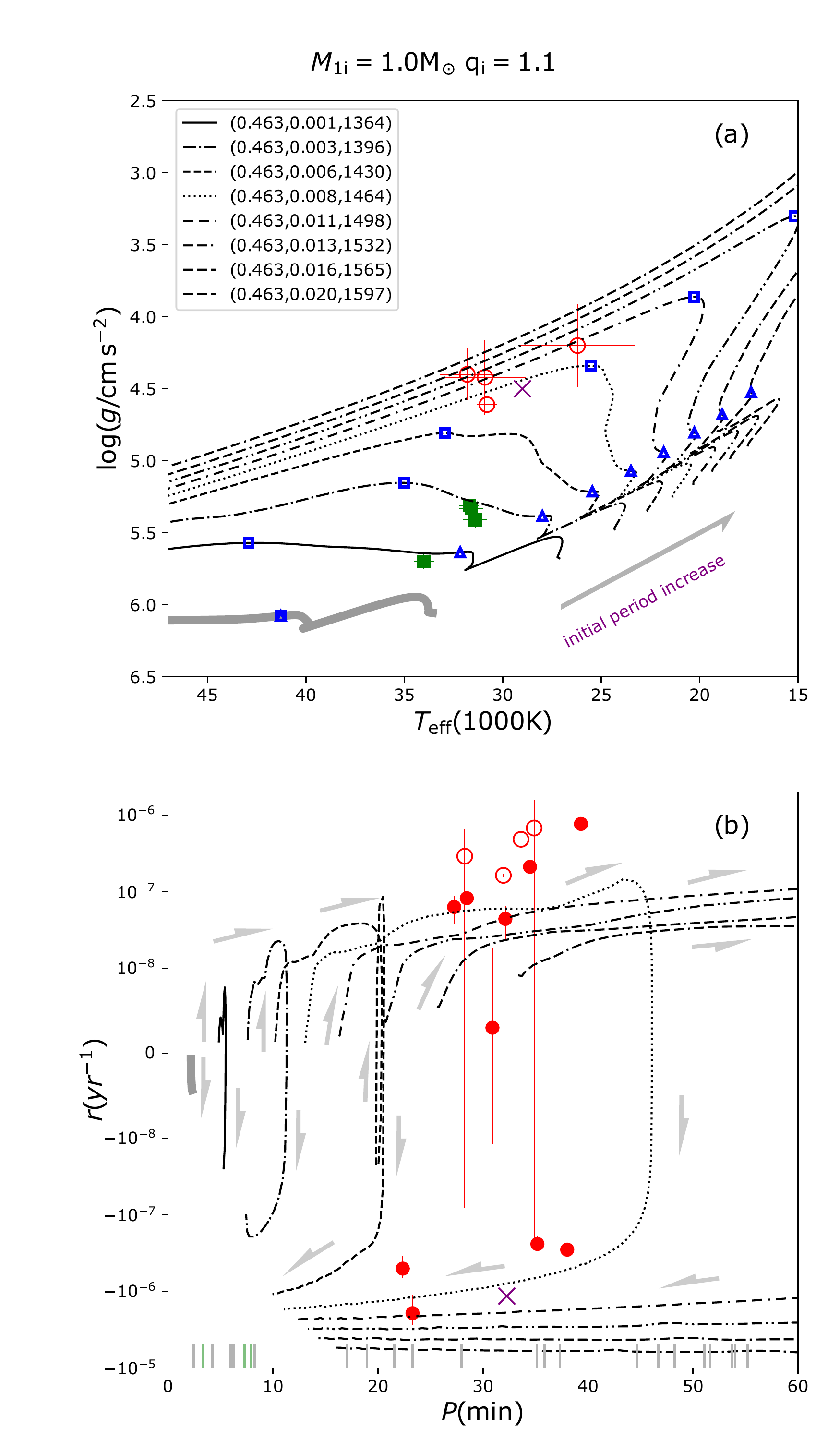}
\caption{Evolution of sdBs produced by binaries, which is same as in Fig.\,\ref{fig combine}, but for ($M_{\rm 1i}, q_{\rm i}$)=(1.0$M_\odot$,1.1).
\label{fig A6}}%
\end{figure}

\begin{figure}
\includegraphics[width=8cm]{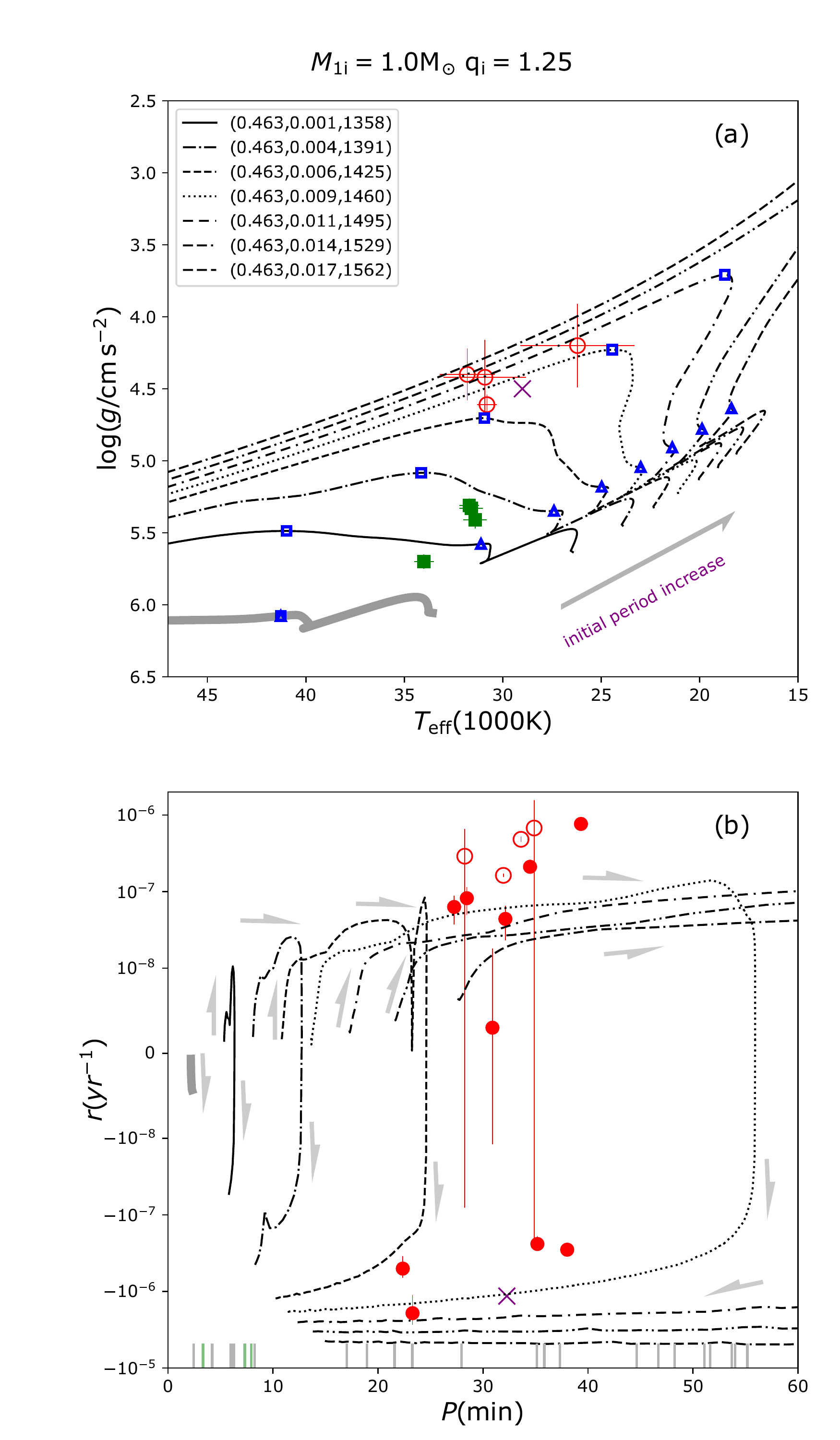}
\caption{Evolution of sdBs produced by binaries, which is same as in Fig.\,\ref{fig combine}, but for ($M_{\rm 1i}, q_{\rm i}$)=(1.0$M_\odot$,1.25).
\label{fig A7}}%
\end{figure}

\begin{figure}
\includegraphics[width=8cm]{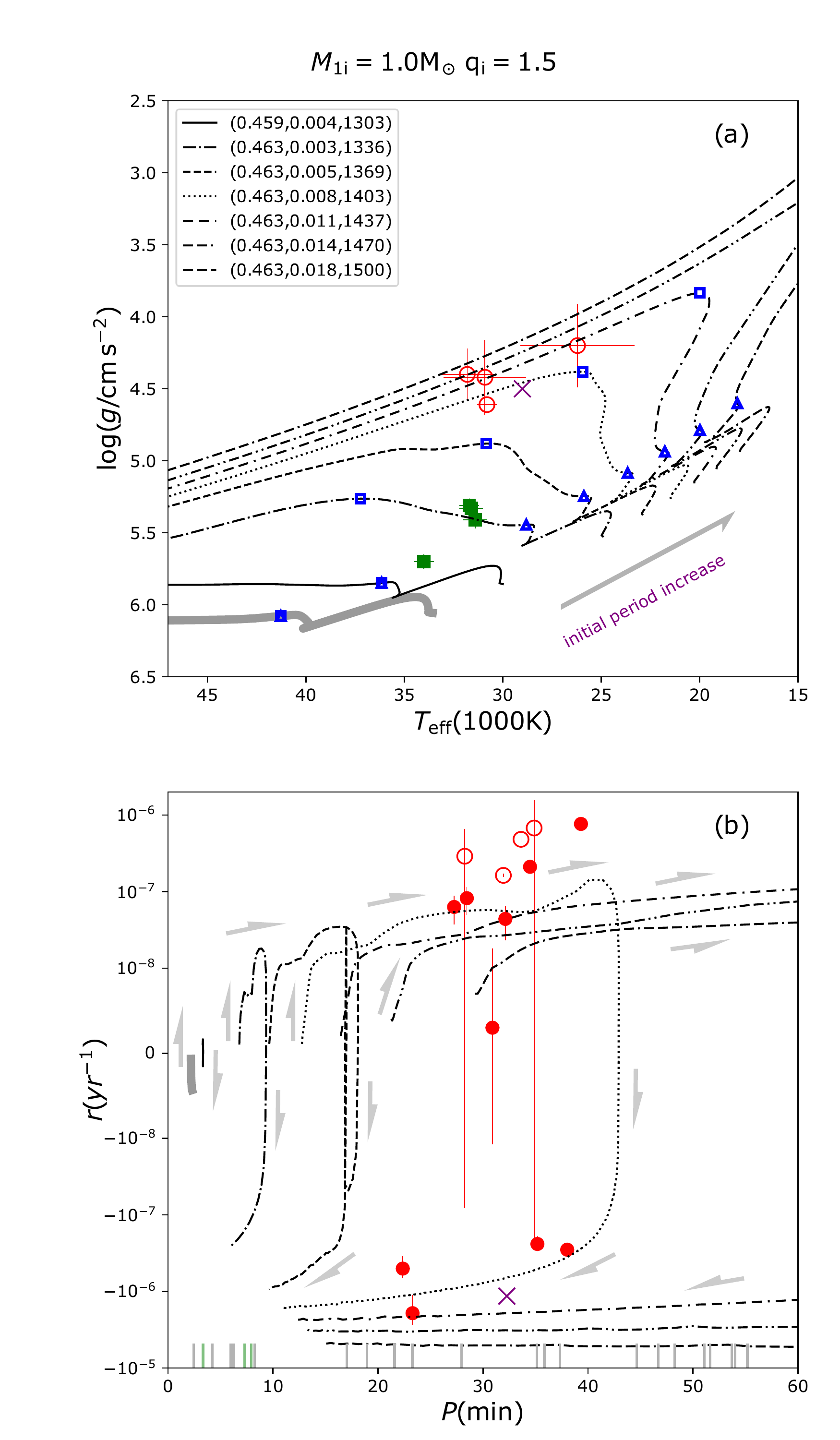}
\caption{Evolution of sdBs produced by binaries, which is same as in Fig.\,\ref{fig combine}, but for ($M_{\rm 1i}, q_{\rm i}$)=(1.0$M_\odot$,1.5).
\label{fig A8}}%
\end{figure}

\begin{figure}
\includegraphics[width=8cm]{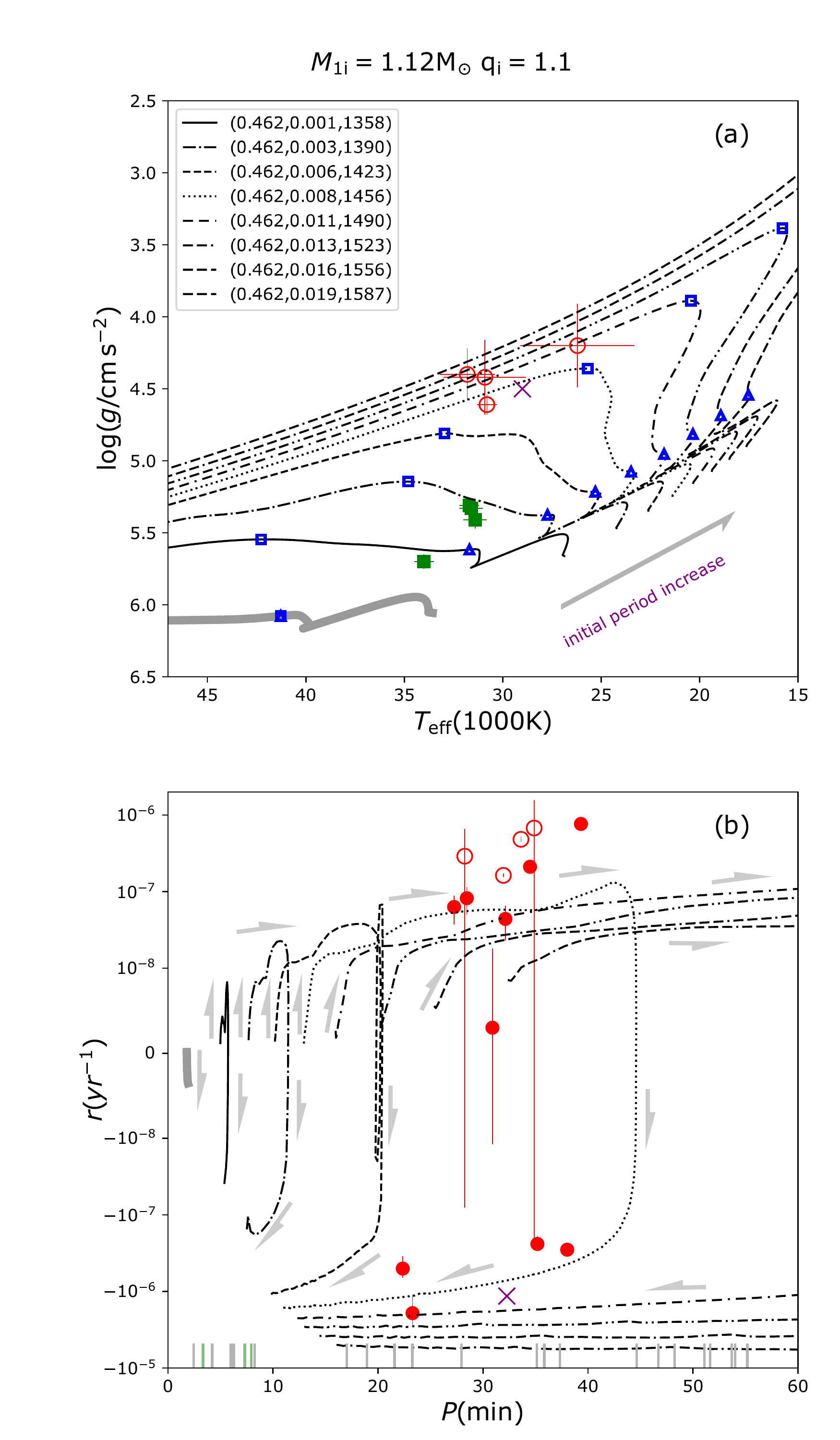}
\caption{Evolution of sdBs produced by binaries, which is same as in Fig.\,\ref{fig combine}, but for ($M_{\rm 1i}, q_{\rm i}$)=(1.12$M_\odot$,1.1).
\label{fig A9}}%
\end{figure}


\begin{figure}
\includegraphics[width=8cm]{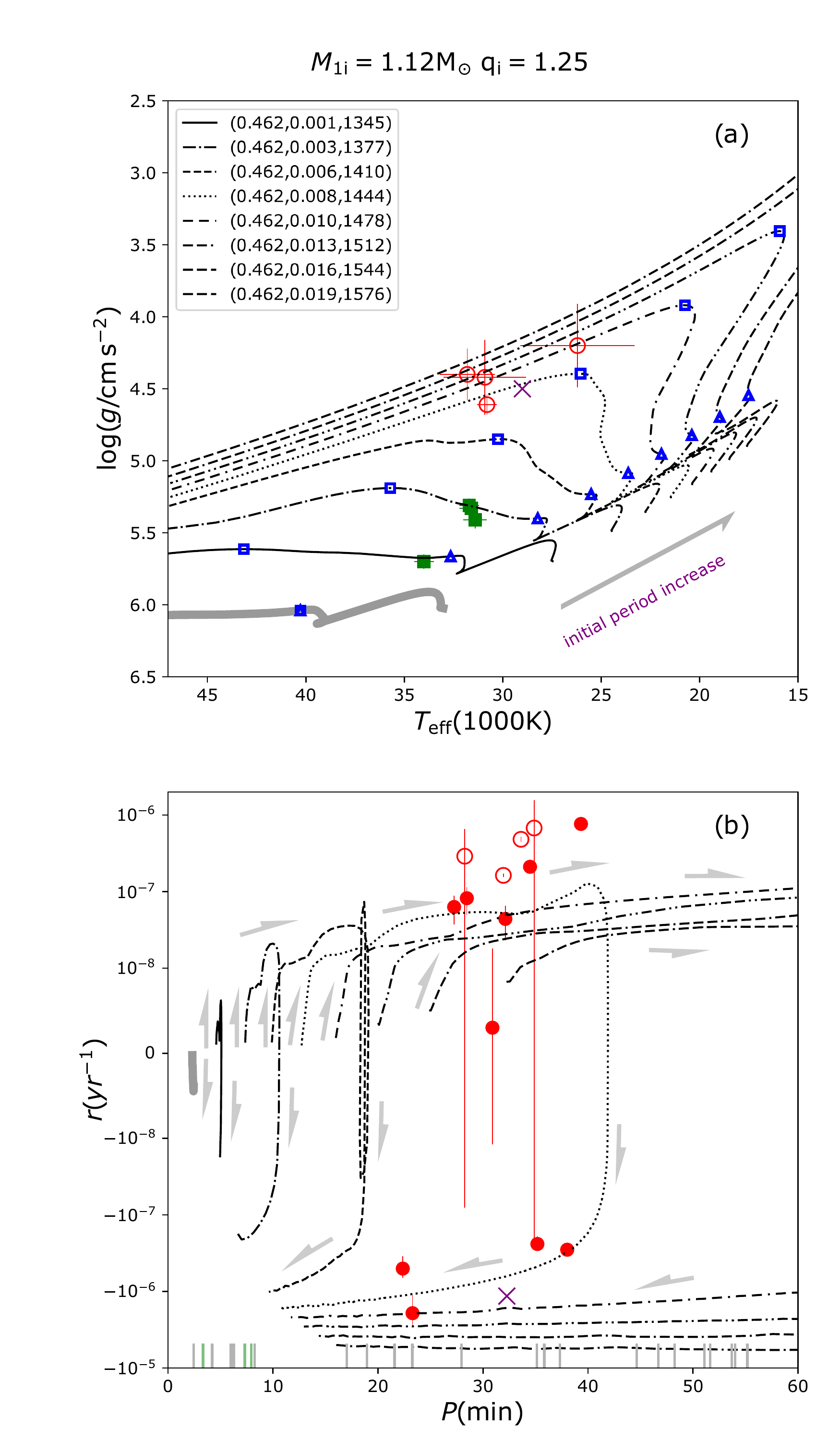}
\caption{Evolution of sdBs produced by binaries, which is same as in Fig.\,\ref{fig combine}, but for ($M_{\rm 1i}, q_{\rm i}$)=(1.12$M_\odot$,1.25).
\label{fig A10}}%
\end{figure}
\clearpage

\begin{figure}
\includegraphics[width=8cm]{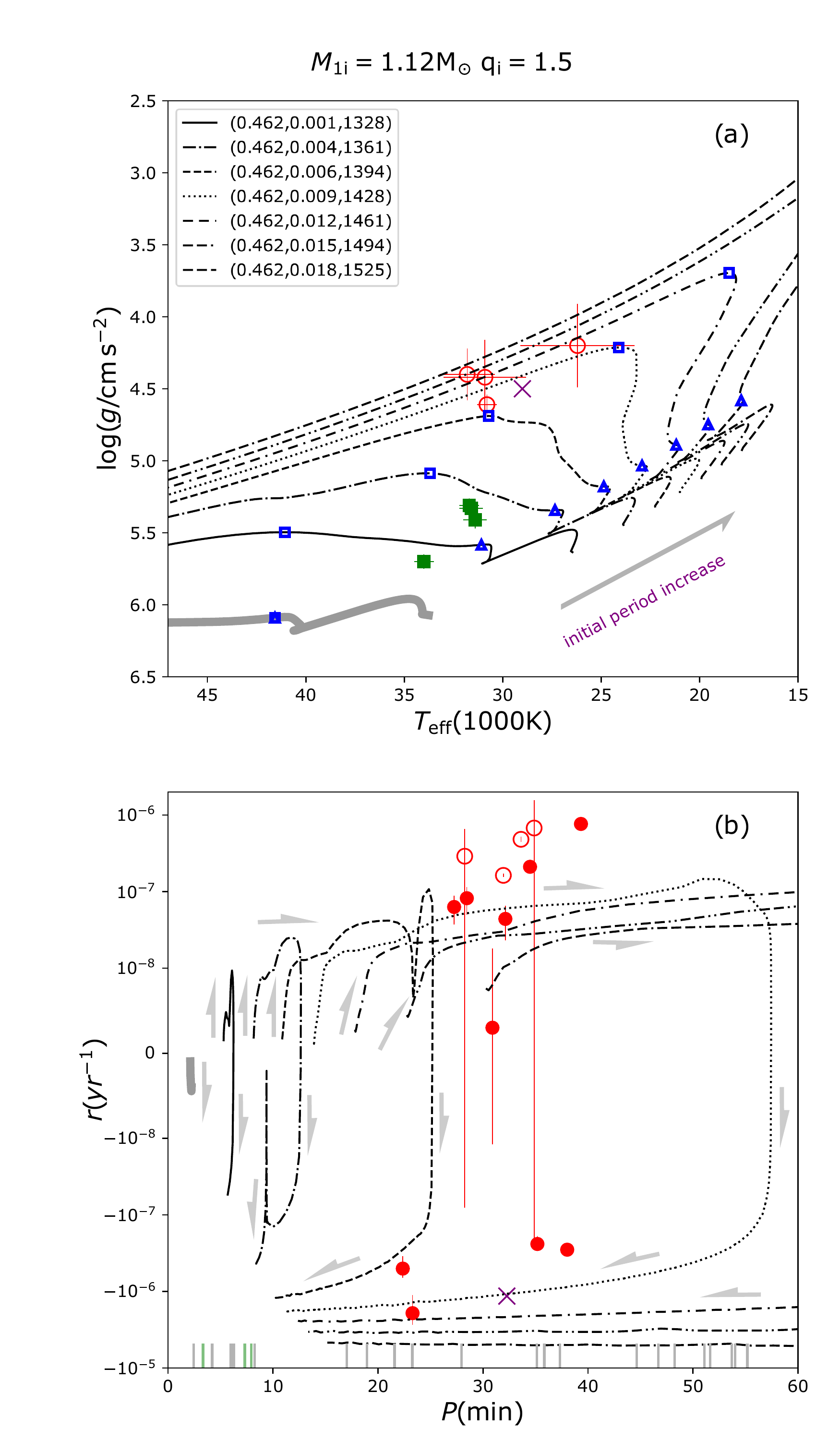}
\caption{Evolution of sdBs produced by binaries, which is same as in Fig.\,\ref{fig combine}, but for ($M_{\rm 1i}, q_{\rm i}$)=(1.12$M_\odot$,1.5).
\label{fig A11}}%
\end{figure}

\begin{figure}
\includegraphics[width=8cm]{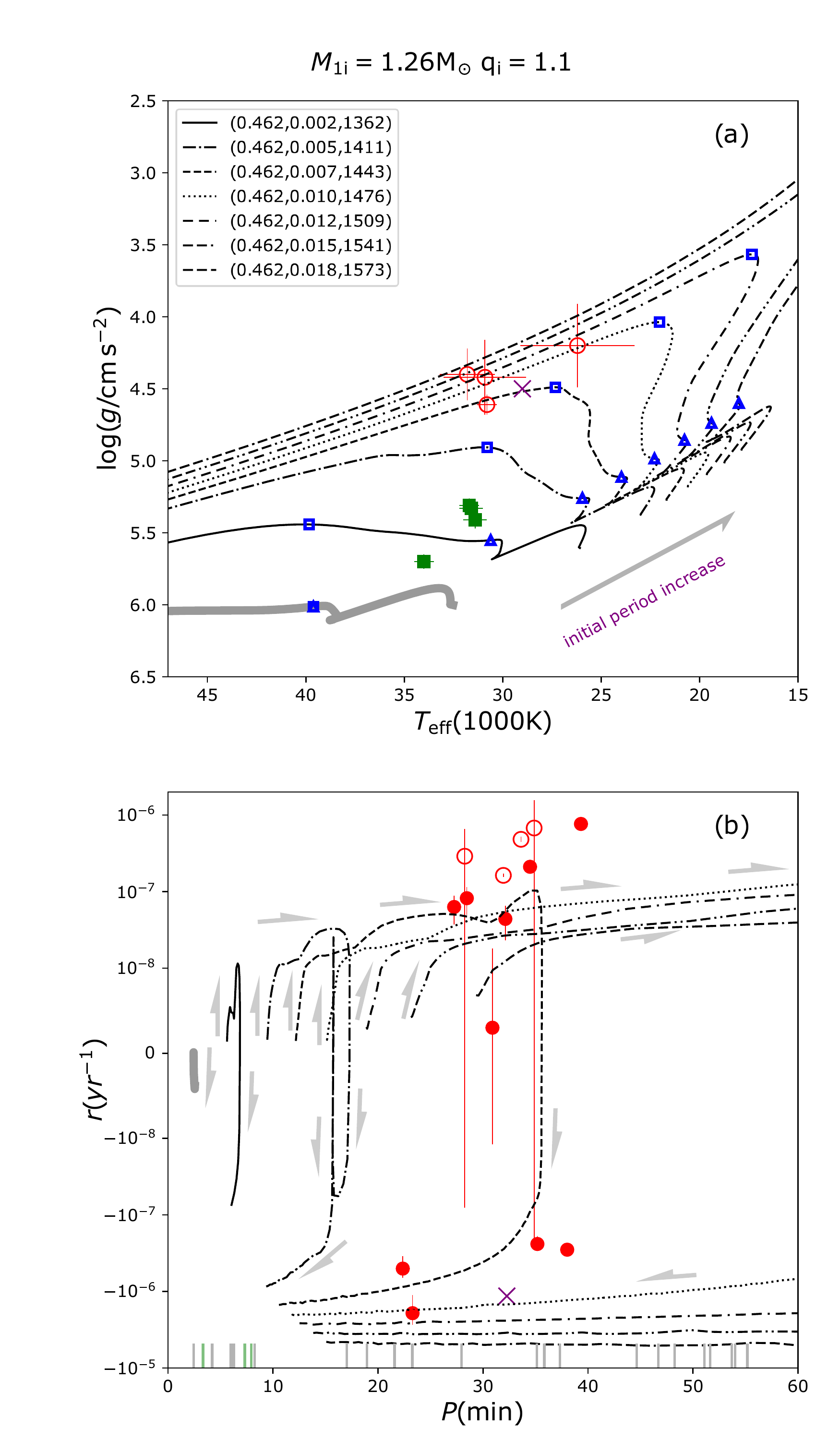}
\caption{Evolution of sdBs produced by binaries, which is same as in Fig.\,\ref{fig combine}, but for ($M_{\rm 1i}, q_{\rm i}$)=(1.26$M_\odot$,1.1).
\label{fig A12}}%
\end{figure}

\begin{figure}
\includegraphics[width=8cm]{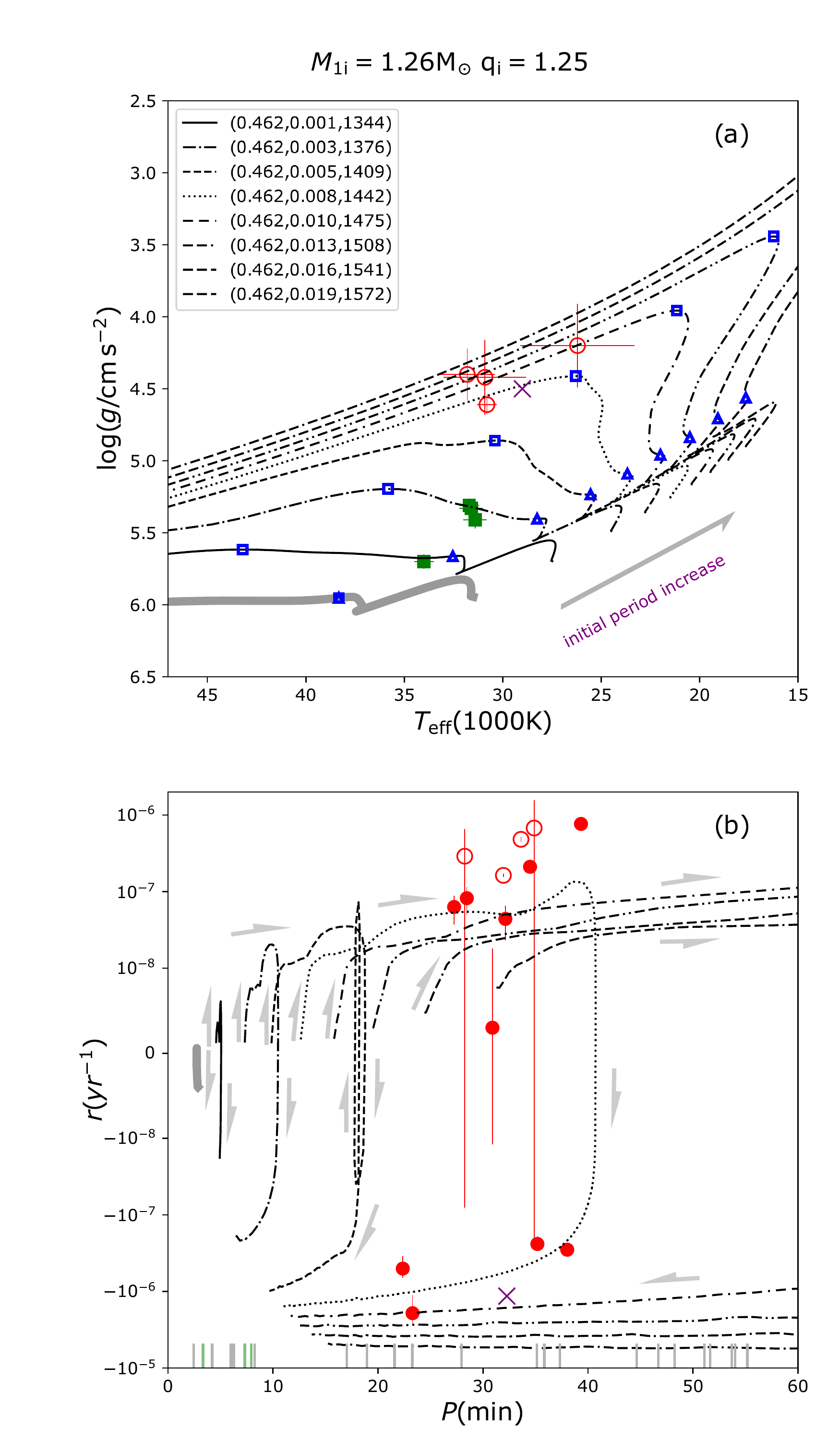}
\caption{Evolution of sdBs produced by binaries, which is same as in Fig.\,\ref{fig combine}, but for ($M_{\rm 1i}, q_{\rm i}$)=(1.26$M_\odot$,1.25).
\label{fig A13}}%
\end{figure}

\begin{figure}
\includegraphics[width=8cm]{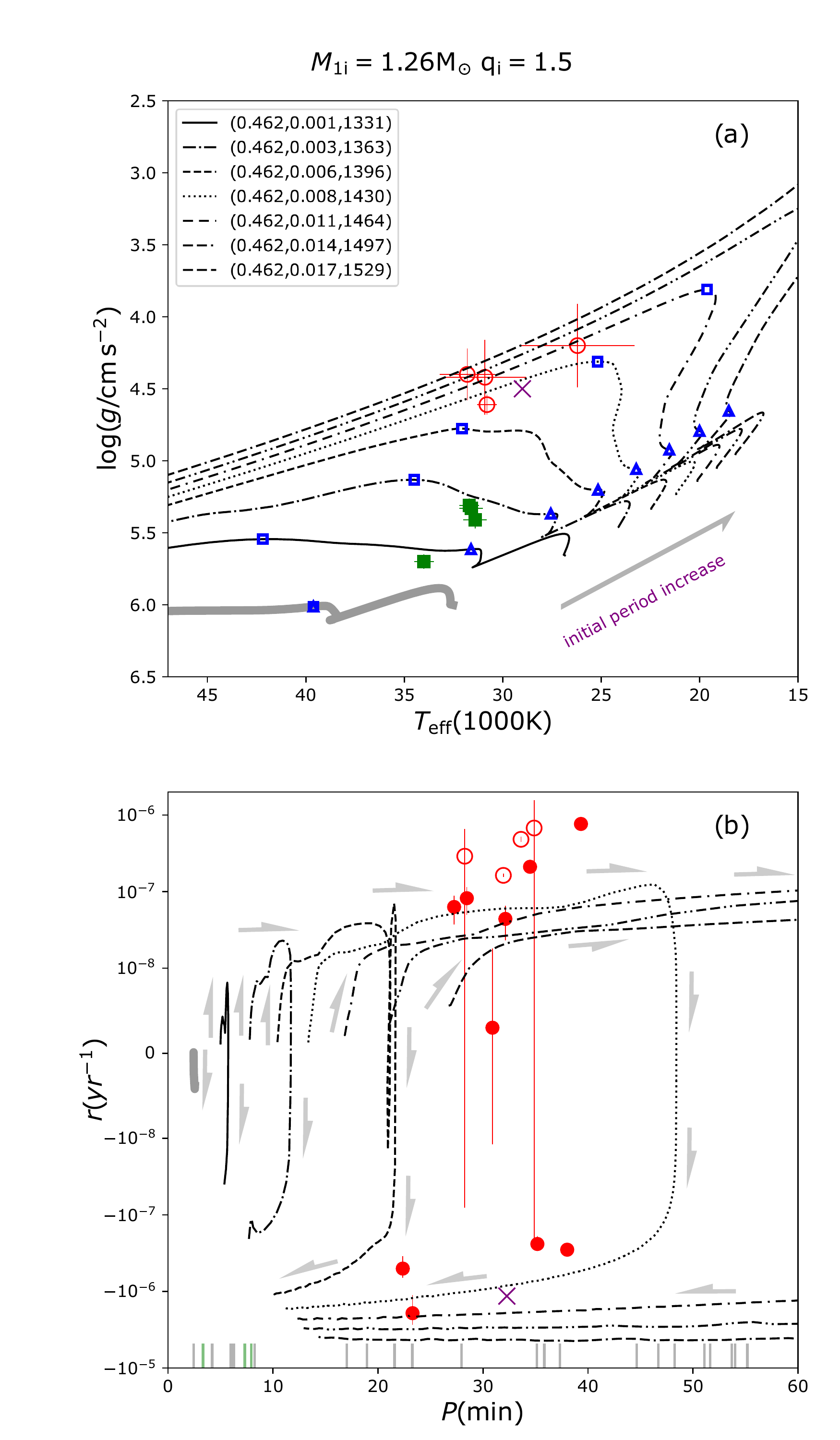}
\caption{Evolution of sdBs produced by binaries, which is same as in Fig.\,\ref{fig combine}, but for ($M_{\rm 1i}, q_{\rm i}$)=(1.26$M_\odot$,1.5).
\label{fig A14}}%
\end{figure}

\begin{figure}
\includegraphics[width=8cm]{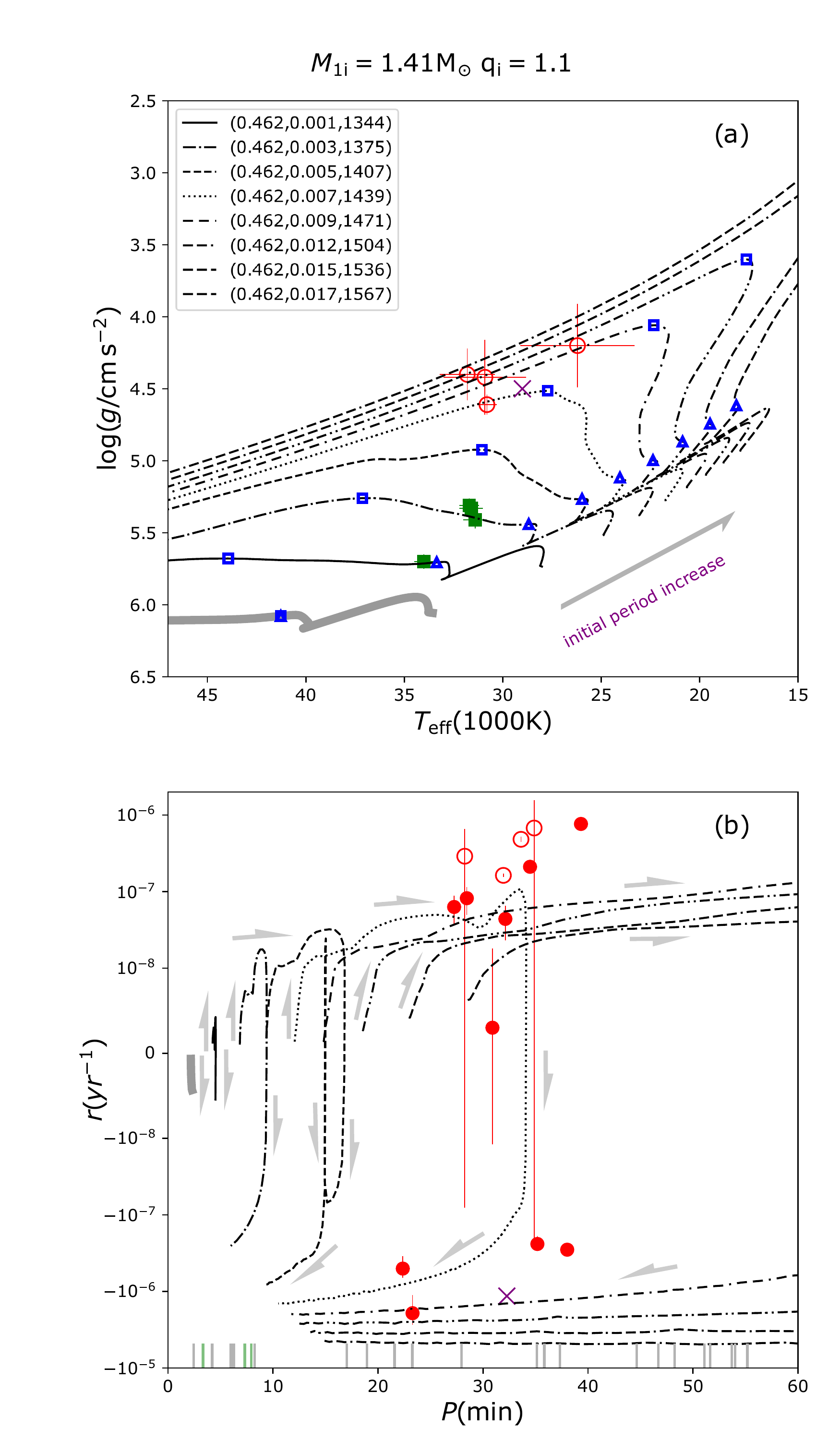}
\caption{Evolution of sdBs produced by binaries, which is same as in Fig.\,\ref{fig combine}, but for ($M_{\rm 1i}, q_{\rm i}$)=(1.41$M_\odot$,1.1).
\label{fig A15}}%
\end{figure}

\begin{figure}
\includegraphics[width=8cm]{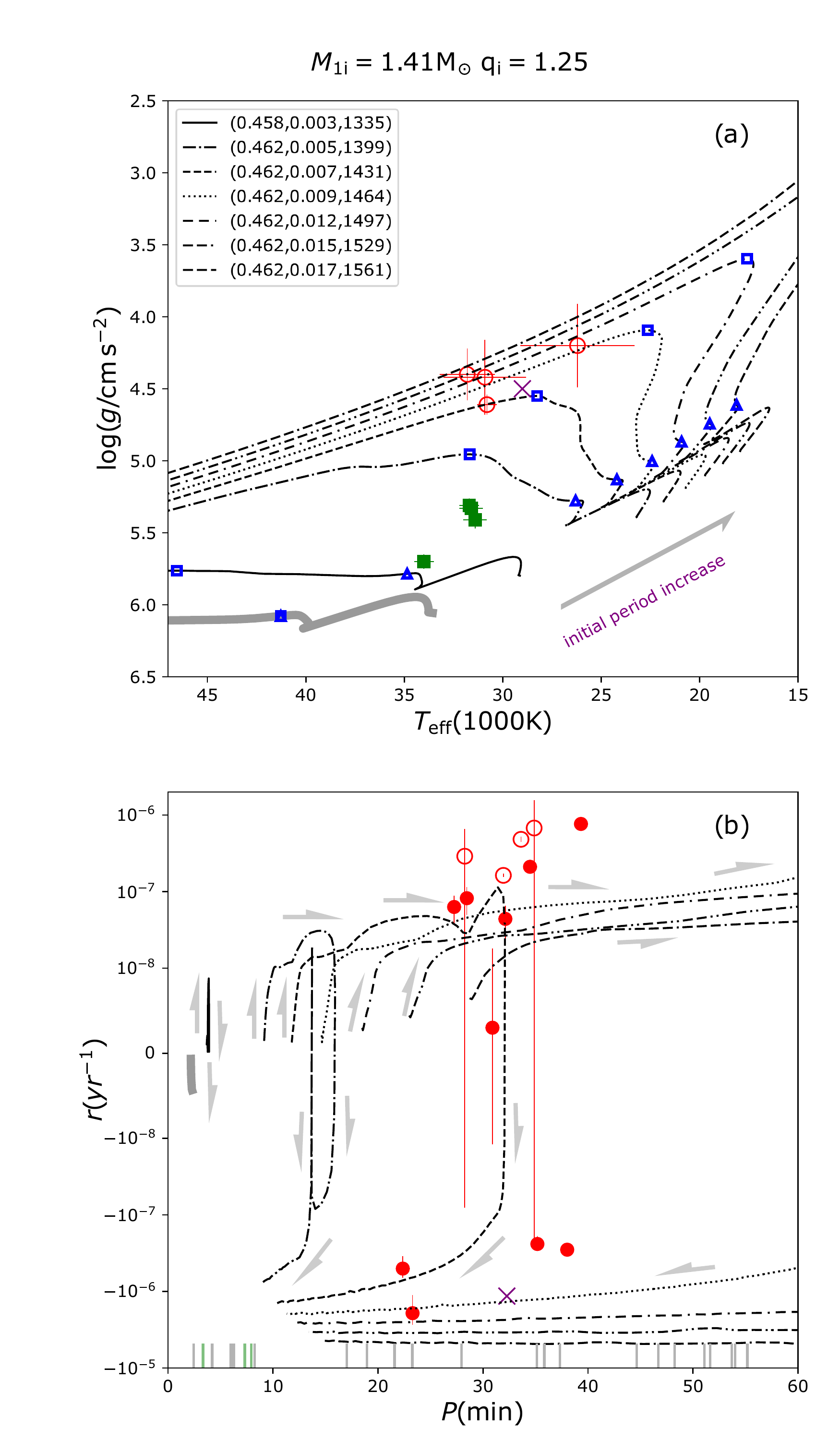}
\caption{Evolution of sdBs produced by binaries, which is same as in Fig.\,\ref{fig combine}, but for ($M_{\rm 1i}, q_{\rm i}$)=(1.41$M_\odot$,1.25).
\label{fig A16}}%
\end{figure}

\begin{figure}
\includegraphics[width=8cm]{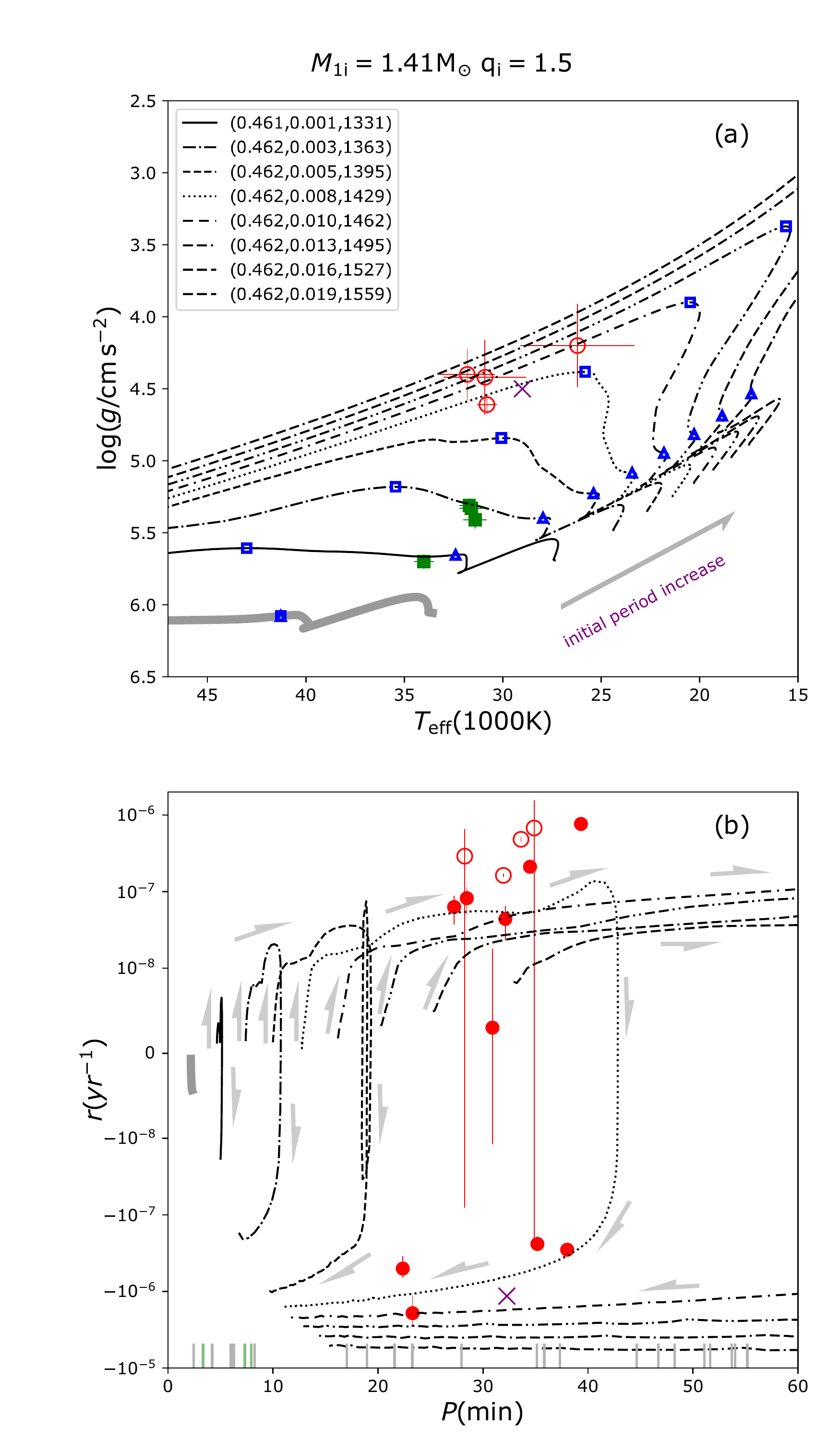}
\caption{Evolution of sdBs produced by binaries, which is same as in Fig.\,\ref{fig combine}, but for ($M_{\rm 1i}, q_{\rm i}$)=(1.41$M_\odot$,1.5).
\label{fig A17}}%
\end{figure}

\begin{figure}
\includegraphics[width=8cm]{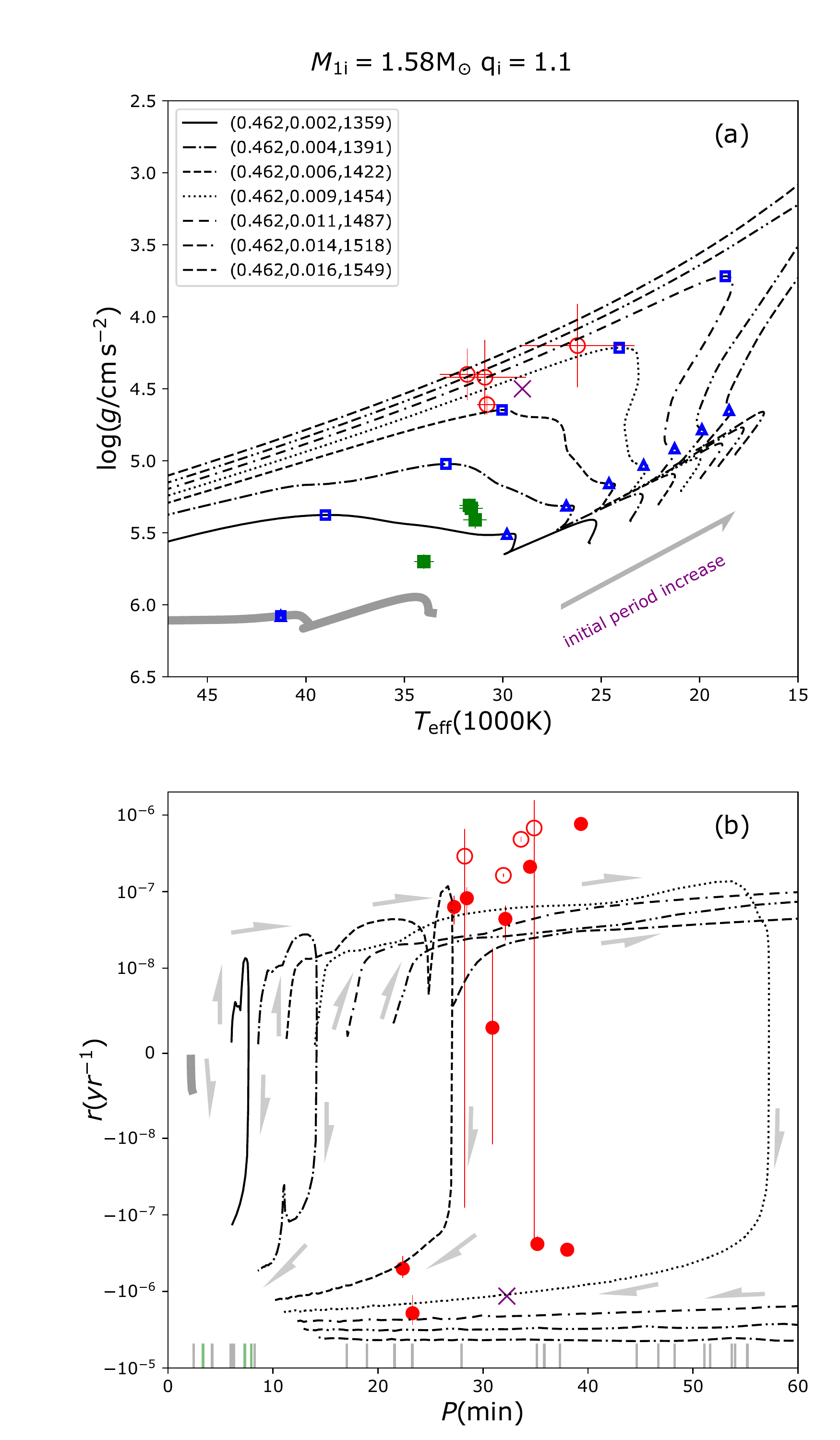}
\caption{Evolution of sdBs produced by binaries, which is same as in Fig.\,\ref{fig combine}, but for ($M_{\rm 1i}, q_{\rm i}$)=(1.58$M_\odot$,1.1).
\label{fig A18}}%
\end{figure}

\begin{figure}
\includegraphics[width=8cm]{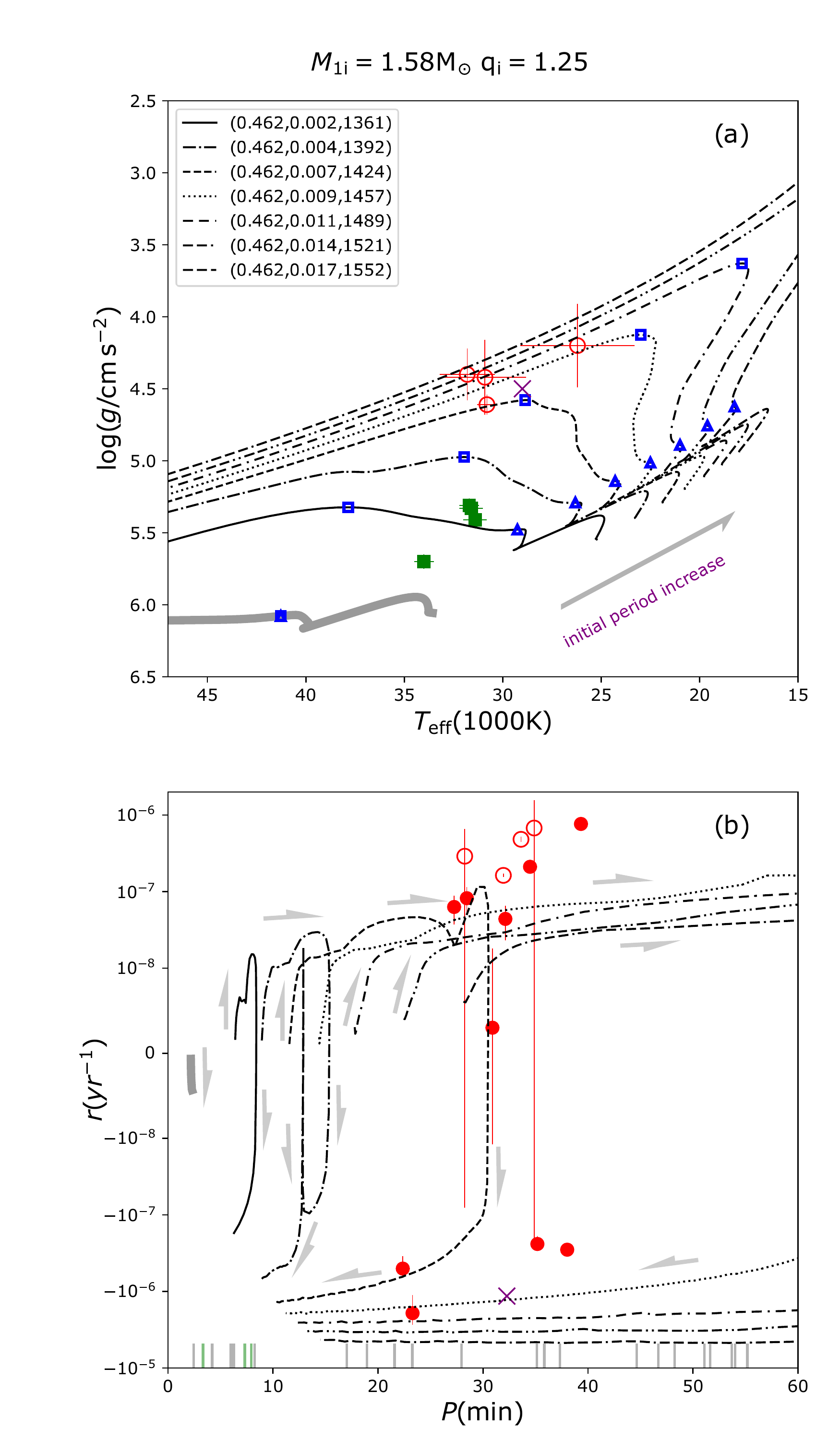}
\caption{Evolution of sdBs produced by binaries, which is same to Fig.\,\ref{fig combine}, but for ($M_{\rm 1i}, q_{\rm i}$)=(1.58$M_\odot$,1.25).
\label{fig A19}}%
\end{figure}

\begin{figure}
\includegraphics[width=8cm]{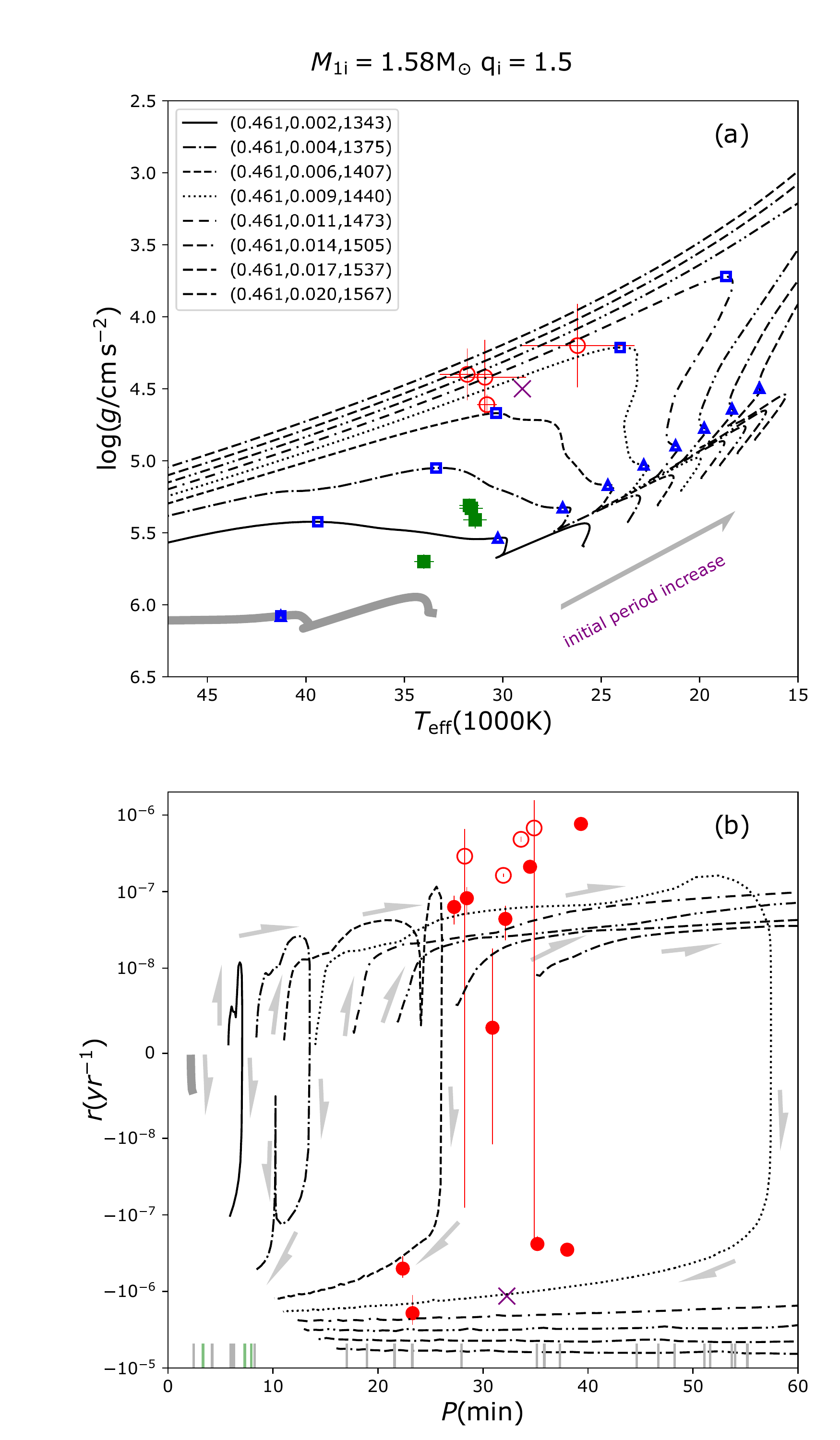}
\caption{Evolution of sdBs produced by binaries, which is same as in Fig.\,\ref{fig combine}, but for ($M_{\rm 1i}, q_{\rm i}$)=(1.58$M_\odot$,1.5).
\label{fig A20}}%
\end{figure}

\begin{figure}
\includegraphics[width=8cm]{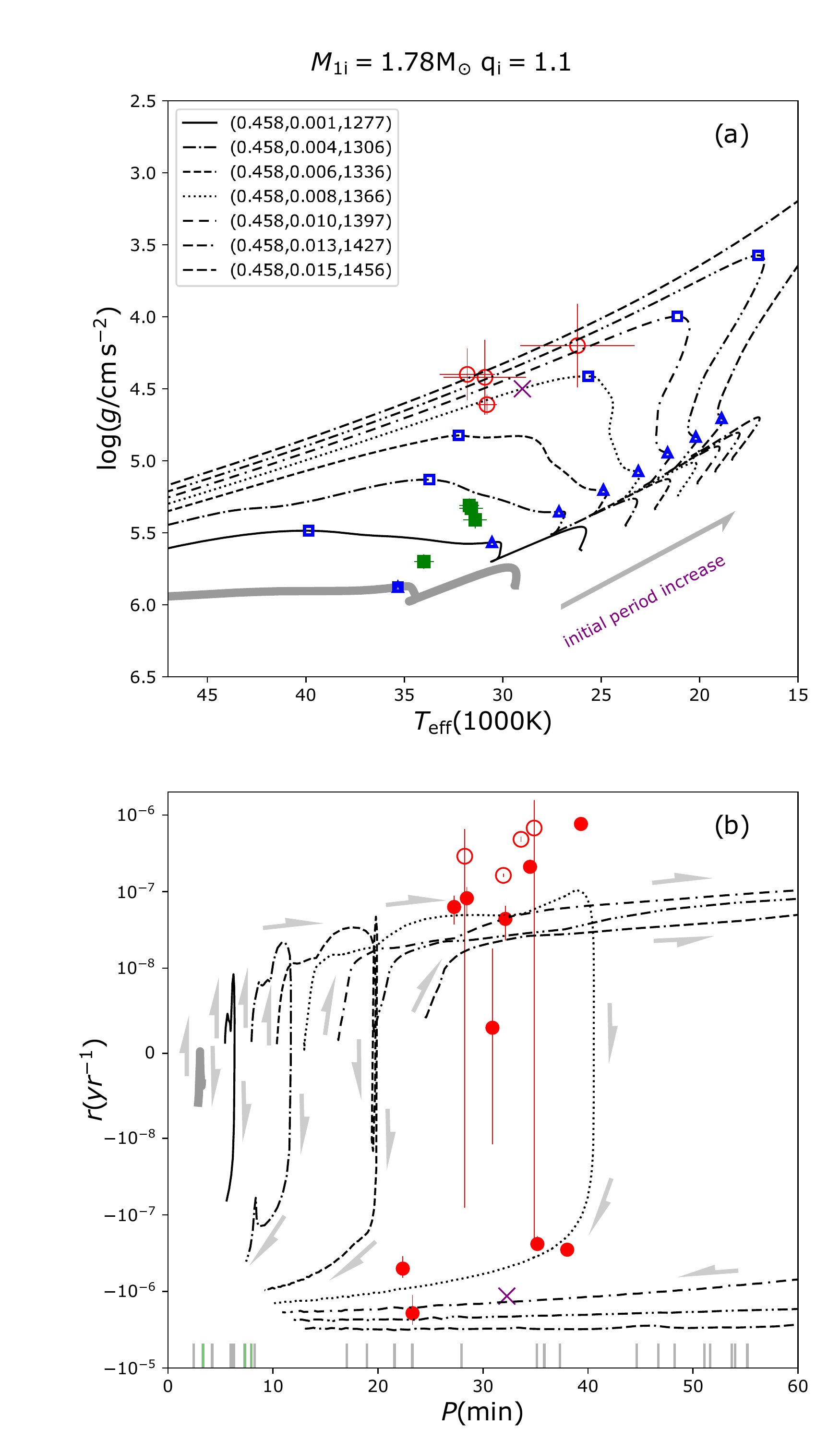}
\caption{Evolution of sdBs produced by binaries, which is same to Fig.\,\ref{fig combine}, but for ($M_{\rm 1i}, q_{\rm i}$)=(1.78$M_\odot$,1.1).
\label{fig A21}}%
\end{figure}

\begin{figure}
\includegraphics[width=8cm]{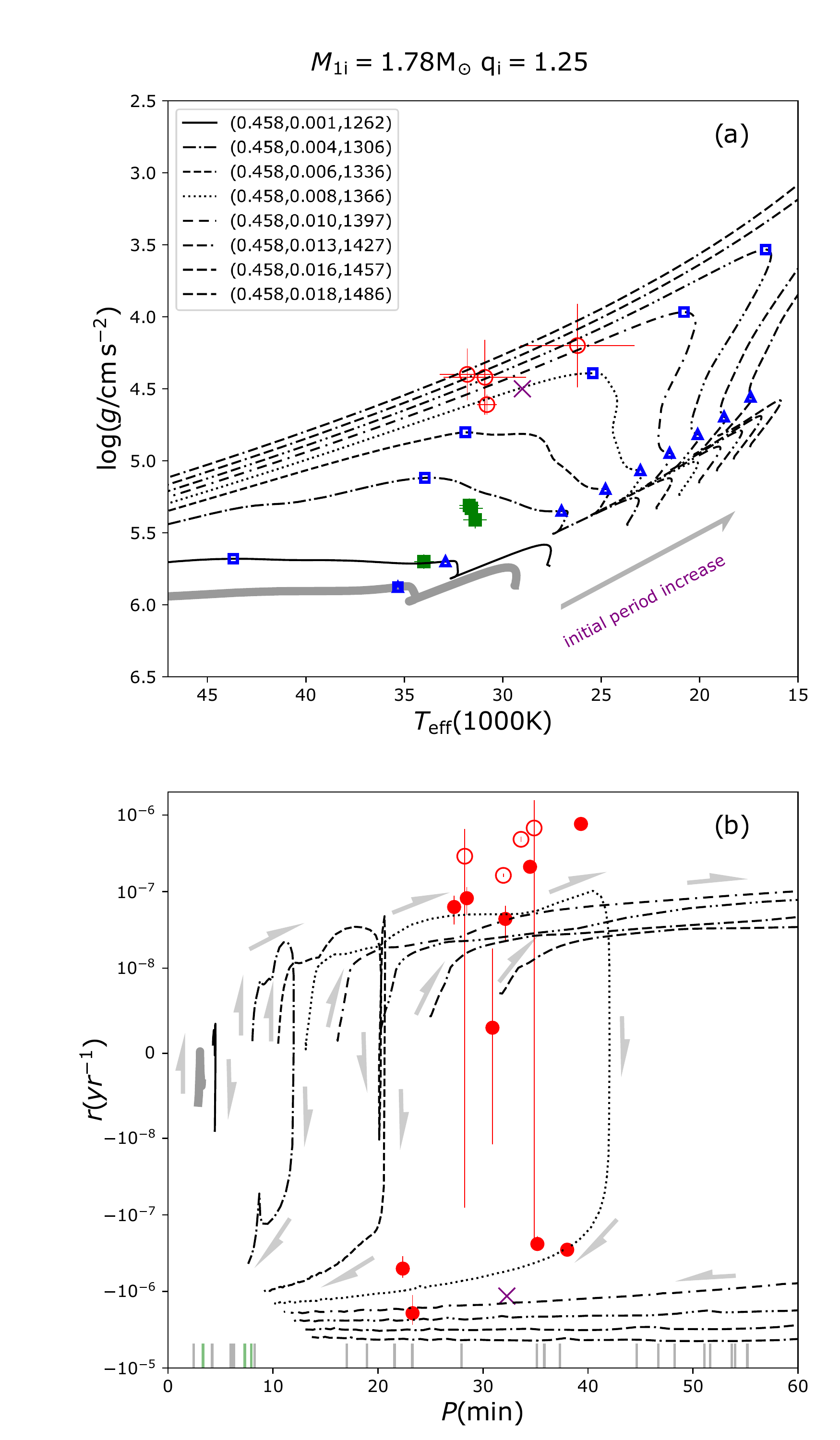}
\caption{Evolution of sdBs produced by binaries, which is same as in Fig.\,\ref{fig combine}, but for ($M_{\rm 1i}, q_{\rm i}$)=(1.78$M_\odot$,1.25).
\label{fig A22}}%
\end{figure}

\begin{figure}
\includegraphics[width=8cm]{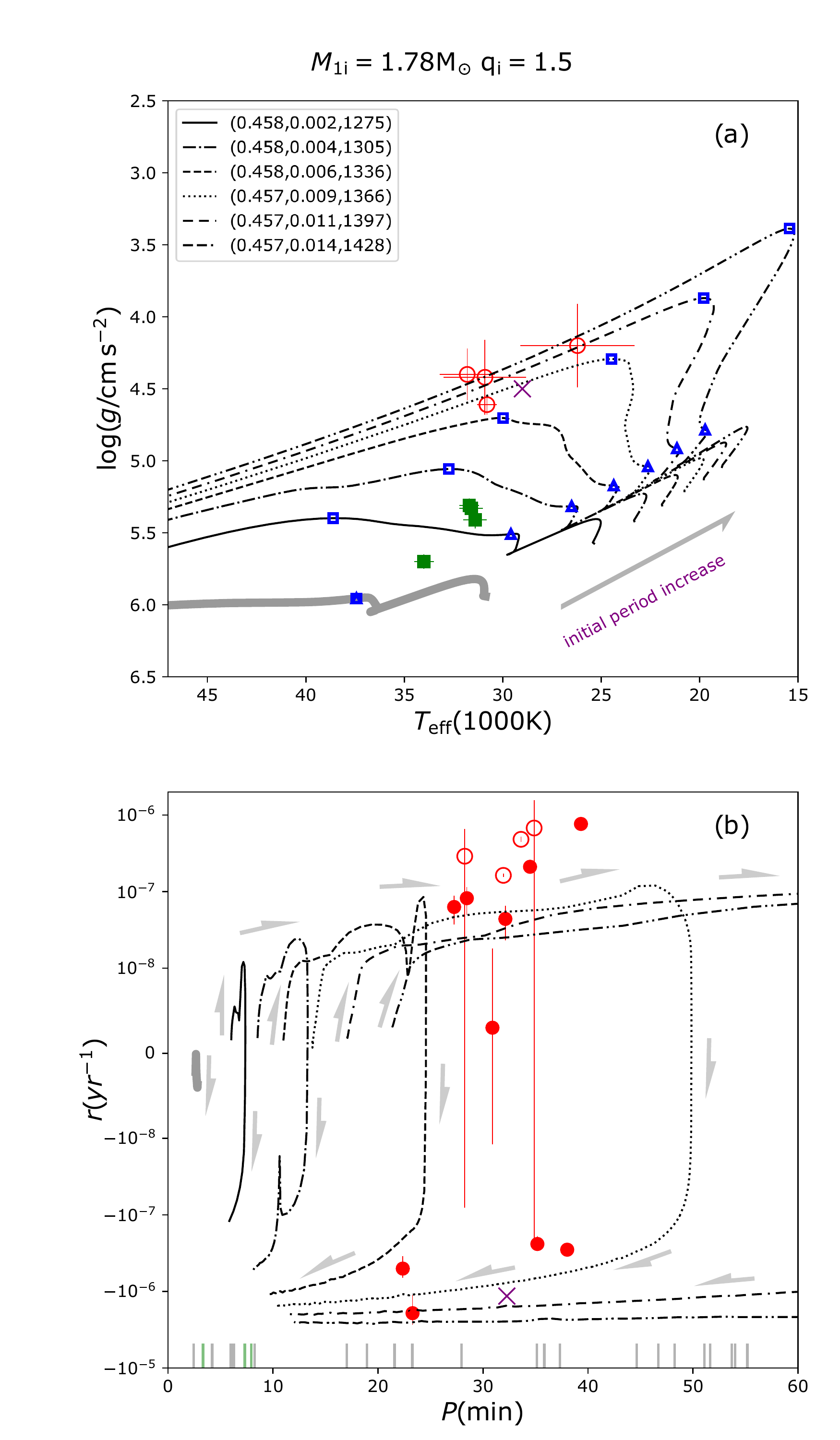}
\caption{Evolution of sdBs produced by binaries, which is same as in Fig.\,\ref{fig combine}, but for ($M_{\rm 1i}, q_{\rm i}$)=(1.78$M_\odot$,1.19).
\label{fig A23}}%
\end{figure}

\end{appendix}
\end{document}